\documentclass[nonacm,acmsmall]{acmart}
\usepackage{algorithm2e}
\usepackage{longtable}
\usepackage{siunitx}
\usepackage{rotating}
\usepackage{subcaption}
\usepackage{hyphenat}

\begin{document}


\title[Feasibility of Large Transformers on FPGA Clusters]{The Feasibility of Implementing Large-Scale Transformers on Multi-FPGA Platforms}

\author{Yu Gao}
\email{yujohn.gao@mail.utoronto.ca}
\orcid{0009-0002-4392-8442}
\affiliation{%
  \institution{University of Toronto}
  \country{Canada}
}

\author{Juan Camilo Vega}
\email{camilo.vega@mail.utoronto.ca}
\affiliation{%
  \institution{University of Toronto}
  \country{Canada}
}

\author{Paul Chow}
\email{pc@eecg.toronto.edu}
\orcid{0000-0002-0523-7117}
\affiliation{%
  \institution{University of Toronto}
  \country{Canada}
}


\begin{abstract}
FPGAs are rarely mentioned when discussing the implementation of large machine learning applications, such as Large Language Models (LLMs), in the data center.
There has been much evidence showing that single FPGAs can be competitive with GPUs in performance for some computations, especially for low latency, and often much more efficient when power is considered.
This suggests that there is merit to exploring the use of multiple FPGAs for large machine learning applications.

The challenge with using multiple FPGAs is that there is no commonly-accepted flow for developing and deploying multi-FPGA applications, i.e., there are no tools to describe a large application, map it to multiple FPGAs and then deploy the application on a multi-FPGA platform.

In this paper, we explore the feasibility of implementing large transformers using multiple FPGAs by developing a scalable multi-FPGA platform and some tools to map large applications to the platform.
We validate our approach by designing an efficient multi-FPGA version of the I-BERT transformer and implement one encoder using six FPGAs as a working proof-of-concept to show that our platform and tools work.

Based on our proof-of-concept prototype and the estimations of performance using the latest FPGAs compared to GPUs, we conclude that there can be a place for FPGAs in the world of large machine learning applications.
We demonstrate a promising first step that shows that with the right infrastructure and tools it is reasonable to continue to explore the possible benefits of using FPGAs for applications such as LLMs.

\end{abstract}





\maketitle

\section{Introduction}\label{CH-INTRO}

Large-scale applications for machine learning, such as Large Language Models (LLMs), are run in data centers on clusters of GPUs~\cite{gpucluster:sc21} or ASIC accelerators~\cite{tpu:isca2023} customized for such applications.
There are no known implementations using FPGAs to run such models at similar scales.
We hypothesize the reason for not seeing the use of FPGAs is that there are two key missing elements in the FPGA ecosystem: infrastructure and tools.
The large machine learning applications are run on clusters of accelerators, oftentimes thousands of them~\cite{numgpus:nextplatform2023}, and there are few known platforms capable of provisioning thousands of FPGAs.
Microsoft's Catapult~v2~\cite{catapult_v2} and Amazon's F1~\cite{aws.Amazon.2022} are the best publicly known FPGA platforms that may have the capability, but there has been no indication that they have been used for such large applications.
A likely reason is that there are no tools that make it even possible to describe and implement large machine learning models on clusters of FPGAs and then deploy them easily, while for GPUs and other accelerators tooling based on environments, such as Tensorflow~\cite{tensorflow}, have been developed~\cite{tpuv1:cacm2018}.

The reasons to consider FPGAs for large-scale machine learning is for the results already demonstrated by FPGAs at much smaller scales.
Many FPGA studies, especially in machine learning, have shown that FPGAs can come close to the performance of GPUs and very often do better when power is also taken into consideration.
We only cite a few examples here~\cite{dlau,fpdnn:fccm2017,dnnbuilder:iccad2018}.
Another benefit of using FPGAs is the ability to customize the memory system and to implement high degrees of pipelining.
This can realize solutions with much lower latency when compared to GPUs~\cite{flexcnn:trets2023}
that require large batch sizes to maximize the benefit of being able to execute multiple threads in parallel.
 
While the definitive proof that any design can work and be beneficial is to build a working system, the scale of implementing large machine learning models means that it would be very high risk to simply try and build a large model on a multi-FPGA platform.
Significant hardware, software and manpower resources would be required.
Before attempting any complex design, it is prudent to determine whether there is merit to even attempting it.

In this paper, our first high-level goal is to demonstrate that with better infrastructure and tools it will be feasible to implement large-scale machine learning applications using FPGAs.
By this we mean that it should be possible to describe the desired machine learning model and then in an automated way turn it into the circuits that can then be deployed on a multi-FPGA platform.
We want to keep the model description at a relatively high level of abstraction so that developers can focus on the functionality and not worry about the low-level implementation details, such as how communication is done.
We achieve our first goal by building a working proof-of-concept, six-FPGA implementation of an encoder layer for I-BERT~\cite{ibert} using our flow and running on our multi-FPGA testbed..
This demonstrates that our platform architecture and tool flow are able to generate the FPGA bitstreams for a fully functional system that could run in a data center environment. 

The second high-level goal of this paper is to use our proof-of-concept platform to judge whether multi-FPGA platforms can compete with clusters of high-end GPUs.
We do comparisons of our current platform, which uses AMD UltraScale+ FPGAs~\cite{sidewinder}, against some current GPUs, the NVIDIA T4~\cite{nv-t4} and the NVIDIA A100~\cite{nv-a100}.
Given that our current system does not use the most recent generation of FPGAs, it is not surprising that our performance does not compete.
However, based on our working proof-of-concept implementation we are able to do a performance estimation assuming the same system is implemented using the latest AMD Versal~\cite{versal} devices.
We show that with the most recent FPGAs we can achieve performance comparable to high-end GPUs for our proof-of-concept application, which at the size of system we built, is essentially is a comparison of modern single FPGAs with single GPUs.
Other works have also seen similar results~\cite{flightllm:fpga2024,zhou:fpga2024}.
Those other works have also shown power advantages of using FPGA over GPUs, which may be the most important benefit of using FPGAs versus GPUs.
Given this, we expect that multi-FPGA applications should be competitive with equivalent multi-GPU implementations, with an expectation of significant power advantages.

With the positive outcomes of our goals, we show that there is promise to continue with exploring the use of multi-FPGA systems to implement large-scale machine learning applications.

More specifically, in this paper
we begin by developing the hardware infrastructure required for deploying large-scale machine learning models on multiple FPGAs.
Our starting point is to adopt an open-source framework called Galapagos~\cite{galapagos,tarafdar:micro2018}.
The Galapagos hardware/software stack is a framework that allows multiple FPGAs and CPUs to communicate directly using standard networking protocols. 
The user describes an application as a graph of streaming compute kernels and the Galapagos flow will partition the kernels across multiple FPGAs and CPUs.

Even though Galapagos provides an excellent heterogeneous platform for users to program both CPUs and FPGAs in an efficient way, it has a hard limit of 256 on the number of kernels that a Galapagos cluster can accommodate, as overhead would significantly impact performance beyond this point. 
A Galapagos cluster is defined as a group of several kernels, where all the kernels can communicate directly with each other using kernel IDs. The current limit is appropriate for small-scale to medium-scale applications. However, larger applications, such as LLMs, could have thousands of kernels, which is beyond the limit of the current Galapagos framework. In this paper, we show how to overcome this limitation by building clusters of Galapagos clusters.

As the number of kernels increases, it is essential to manage the intra-cluster and inter-cluster communications more efficiently. We develop a message-passing protocol that fits in the Galapagos framework, called the \textit{Galapagos Messaging Interface} (GMI), which enables intra-cluster and inter-cluster collective communication capabilities.

To allow fast deployment of large-scale machine-learning applications, we design an automation tool called the \textit{Cluster Builder} that uses user-defined configuration files and pre-defined libraries to generate IP cores for each kernel, which are later used by the Galapagos toolchain for generating Galapagos clusters. The Cluster Builder also utilizes the GMI functionalities for handling complex communications in the applications.

To demonstrate the scalability and efficiency of the extended Galapagos framework, we accelerate the inference phase of an integer-quantized BERT-base model, called Integer-BERT (I-BERT)~\cite{ibert}, and use the Cluster Builder to deploy the model on multiple Galapagos clusters.

The current Galapagos framework does not support the AMD Versal ACAP~\cite{versal} architecture, thus, we propose a modified Galapagos framework that enables the Galapagos kernels to utilize the AI Engines (AIEs) available in Versal. Based on the architecture of the modified Galapagos framework and the performance of the BERT-base model deployed on the Sidewinder Ultrascale+ FPGAs~\cite{sidewinder,zu19eg}, we estimate the model performance when using the latest AMD Versal devices~\cite{versal} to accelerate it. 

The emphasis of this paper is to define the capabilities required in the infrastructure and tools sufficient enough to demonstrate functionality with reasonable performance.
There are many opportunities to explore more optimal solutions, but that is left for future work.
Again, the focus of this paper is to explore the foundational infrastructure required to deploy large FPGA-based applications.
This will show that there is enough promise in the approach such that it is worthwhile to continue efforts along this path.
With a viable infrastructure solution, it is then reasonable to explore approaches to optimizing the performance.

In summary, the contributions of this work are as follows:

\begin{itemize}
    \item Developed the enhanced Galapagos framework to allow multiple Galapagos clusters to run simultaneously, which enables scaling to thousands of kernels.
    \item Developed the Galapagos Messaging Interface (GMI) layer on top of the current Galapagos framework for handling intra-cluster and inter-cluster communications.
    \item Developed a Cluster Builder that partitions a machine learning model into multiple Galapagos clusters. The Cluster Builder targets various machine learning models and uses the GMI for handling the non-trivial communication patterns that exist in some machine learning models like transformers.
    \item As a proof-of-concept, we have built an efficient multi-FPGA I-BERT model using \textit{High-Level Synthesis} (HLS) and deployed an encoder layer on six FPGAs as a test case for the cluster builder. Based on the performance of one encoder layer, we estimate the overall performance of the I-BERT model.
    \item Proposed a modified Galapagos framework that supports the Versal ACAP~\cite{versal} architecture and estimated the performance of the I-BERT model  when using multiple Versal devices.
    \item Shown that further work on exploring the use of FPGAs for large-scale machine learning applications is clearly merited, with the potential for significant power advantages at little or no performance loss.
\end{itemize}

Section~\ref{CH-BACKGROUND} provides an introduction to the principles and tools that are utilized throughout the work. Section~\ref{CH-RELATED-WORK} discusses related works. 
Section~\ref{CH-SCALING} begins our presentation of Galapagos enhancements, starting with how scaling to larger numbers of kernels is achieved.
Section~\ref{CH-GMI} gives implementation details of the Galapagos Messaging Interface (GMI), which is how collective communication is achieved.
The Cluster Builder is described in Section~\ref{CH-CLUSTERBUILDER} and our multi-FPGA I-BERT model implementation is presented in Section~\ref{CH-TRANSFORMER}. Section~\ref{CH-RESULT} gives performance results of our multi-FPGA I-BERT implementation. Section~\ref{CH-VERSAL} proposes the modified Galapagos framework that supports the recent Versal architecture and estimates the I-BERT performance using Versal. Section~\ref{CH-CONCLUSION} reflects on what we have achieved so far and how we feel that further exploration of the feasibility of implementing large machine learning models with FPGAs is merited. Section~\ref{CH-FUTURE} discusses future works that further improve the design efficiency and framework generality.
\section{Background}\label{CH-BACKGROUND}
In this section, we present the background knowledge and technologies that are utilized and serve as the foundation of this work.

\subsection{Galapagos}\label{CH-Background-galapagos}

Galapagos~\cite{galapagos,tarafdar:micro2018} is a highly modular heterogeneous deployment stack that provides the ability to build clusters of FPGA and CPU nodes that can run any application that can be described as a graph of streaming compute kernels.
All FPGAs and CPUs are directly connected to the data center network.
The FPGAs appear to the application as an abstraction of one large FPGA fabric onto which the streaming kernels are mapped.
The main inputs to Galapagos are a graph describing the connections between the streaming kernels and the actual kernels described in HLS C++ or RTL.
There is also a mapping file that describes how the kernels are mapped to FPGAs or to CPUs.
The mapping file can be manually written or more likely created by a partitioner that can take as input the sizes of the kernels, the latencies, bandwidths and the available devices and the connectivity~\cite{mazraeli:fpl2023}.
After mapping, each FPGA will have one or more of the streaming kernels.
The Galapagos flow will take the inputs and add all the communication IP required so that the kernels can communicate according to the original graph of connections.
This will include switches for kernels communicating within an FPGA and the routing hardware required for FPGA-to-FPGA communication on an external network.
Finally a bitstream for each FPGA in the Galapagos cluster is output.

By using the \textit{libGalapagos} library~\cite{libgalapagos}
Galapagos can also incorporate kernels running on CPUs that can communicate with exactly the same protocol as is used by kernels communicating strictly in hardware.
This makes it easy to build a heterogeneous CPU and FPGA Galapagos cluster.
If the kernels are written in HLS then \textit{libGalapagos} enables a very powerful debugging and development environment.
A cluster of network-connected FPGA designs  written in HLS can be first run entirely in software using \textit{libGalapagos} where the debugging is much easier.
By changing a parameter in one of the Galapagos mapping files any of the FPGA designs can then be mapped to real hardware while the remaining FPGAs are run in software without changing any of the software code.
After all FPGAs have been individually validated in hardware the full hardware system can be deployed.

\begin{figure}
    \centerline{
    \includegraphics[width=0.3\linewidth]{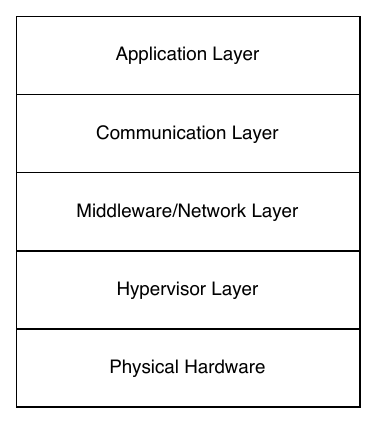}
    }
    \caption{Galapagos hardware stack}
    \label{fig:gp-stack}
\end{figure}

Figure~\ref{fig:gp-stack} shows how
the Galapagos hardware stack is composed of several layers. The Application Layer accommodates the user-provided Galapagos kernels, where the kernels are streaming functions that collectively implement the computations of the application.
The user abstraction is one very large FPGA fabric into which the kernels are mapped and communicate with each other.  The Communication Layer provides an abstraction interface for the kernels and allows the kernels to achieve point-to-point communications within the Galapagos cluster. The Middleware/Network Layer provides the mapping between kernel IDs and FPGA addresses such as MAC and IP addresses and is responsible for sending and receiving messages to and from the correct FPGAs using various network protocols like Ethernet, TCP, or UDP. The Hypervisor~\footnote{Today, this is more commonly called the "Shell" using the terminology from~\cite{catapult_v1}.} Layer provides a static shell for handling I\slash O interfaces and provides standardized communication protocol interfaces to the Middleware/Network Layer. The Physical Hardware Layer represents the actual FPGA hardware without any abstractions. 
The layered approach of Galapagos makes it easier to modify the functionality of Galapagos at one layer without affecting the other layers.

The largest application of Galapagos prior to this work is currently the \textit{AIgean} project~\cite{aigean} described in Section~\ref{CH-Toolchains}.
AIgean sits at the Application Layer of Galapagos as AIgean generates streaming kernels and the inputs to Galapagos that describe the connectivity.

\begin{figure}
    \centerline{
    \includegraphics[width=0.6\linewidth]{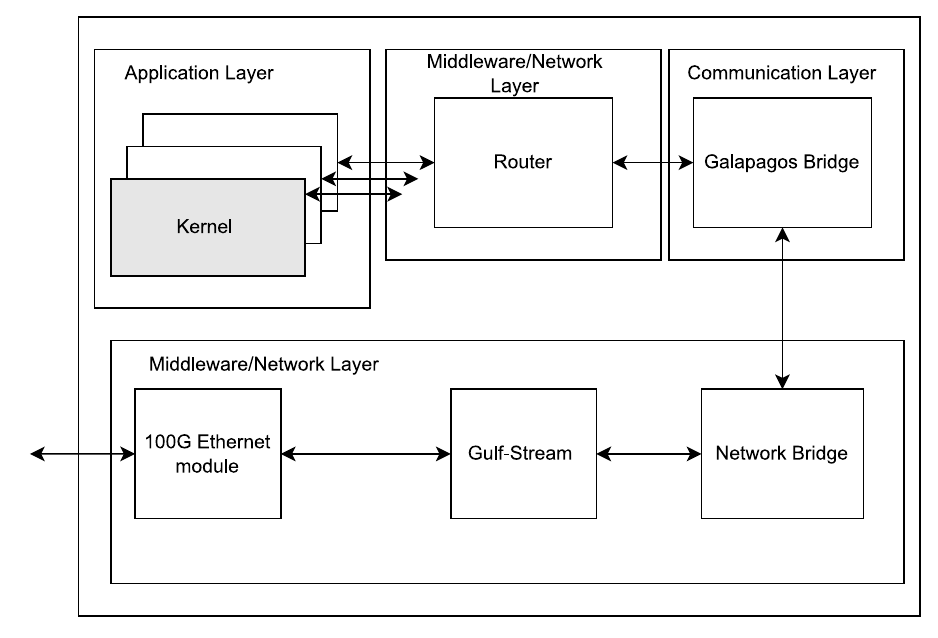}
    }
    \caption{Galapagos hardware architecture}
    \label{fig:gp-arch}
\end{figure}

In this work, we adopt a current version of the Galapagos project that communicates over an IP-based network as the underlying framework. As shown in Figure~\ref{fig:gp-arch},
the hardware architecture consists of a 100G Ethernet module, a UDP core called Gulf-Stream~\cite{gulfstream}, a Network Bridge, a Galapagos Bridge, and a Router. Gulf-Stream provides the UDP connection. The Network Bridge provides the mapping between kernel IDs and FPGA IP addresses. The Router decides whether a packet should be sent to a kernel within the same FPGA or to a kernel in other FPGAs. The Galapagos Bridge provides a packet header that contains information about the sender ID, receiver ID, as well as message size.

Because of the layered structure of Galapagos, it is possible to modify the communication medium and protocol without requiring any changes at the application level.
Our current IP-based system uses a 100~Gbps UDP core.
While UDP is not reliable, it works well-enough in our testbed to build our working proof-of-concept.
We could add reliability using an enhanced UDP such as LTL, which was used in Microsoft's Catapult v2~\cite{catapult_v2}, but we did not feel the effort was necessary for a proof-of-concept that is already working.
We have also run Galapagos using a TCP/IP stack, but prefer the lighter weight UDP core.
In the future, we hope to test our proof-of-concept using a new reliable link-layer protocol called RIFL~\cite{shen2022rifl}.

Galapagos is the foundation of the project presented in this paper.
We use Galapagos as the platform for building clusters of FPGAs.
The baseline Galapagos does not have all the required features to implement very large-scale applications.
An important part of the work presented in this paper is the development of additional features that we have added to Galapagos to support scaling. 
We refer to this modified Galapagos as the \textit{enhanced Galapagos}.

\subsection{Message Passing Interface}\label{CH-Background-mpi}
The Message Passing Interface (MPI)~\cite{mpi} is a common communication protocol for high-performance computing applications. MPI provides a wide range of communication operations including point-to-point operations and collective communication operations. The point-to-point operations allow two processes to communicate with each other directly. The collective communications include operations like Broadcast, Reduce, Gather and Scatter. 

MPI provides a mechanism called the communicator that divides processes or nodes into groups. A group consists of several processes and the communicator is used to assign each process an integer rank for identification. The communicator functions like a handle for identifying groups of processes and providing information on contexts of communication. 

The communicator classifies communication into intra-communicator communications and inter-communicator communications. A process can either use an intra-communicator to accomplish communication within its own group or uses an inter-communicator to accomplish communication across groups. The inter-communicator mechanism also provides a convenient way to dynamically allocate processes when not all processes are pre-allocated at compile time. For instance, when several processes are not created at the start of a program, we can place them in a group that is to be allocated later during the execution of the program.

\subsection{BERT}\label{CH-Background-transformer}

\begin{figure}
    \centerline{
    \includegraphics[width=0.75\linewidth]{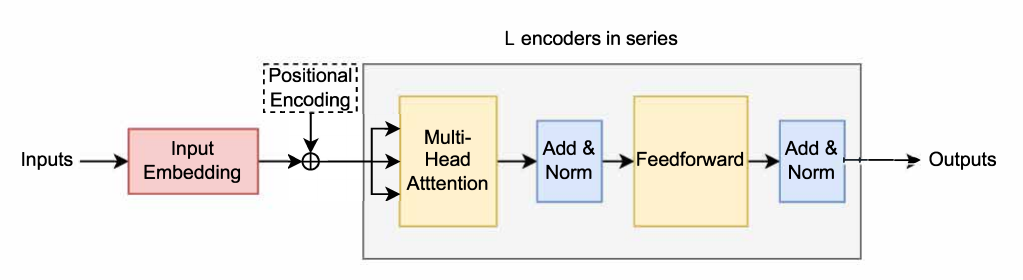}
    }
    \caption{BERT architecture}
    \label{fig:tf-arch}
\end{figure}

As shown in Figure~\ref{fig:tf-arch},
BERT~\cite{bert} is typically composed of an input embedding layer and $L$ encoders. Each encoder mainly consists of a Multi-Head Attention layer, an Add \& Norm layer, a Feedforward layer, and another Add \& Norm layer. The $L$ encoders are connected in series. The BERT structure is described by three parameters: the number of encoders ($L$), the number of attention heads ($A$), and the hidden size ($H$). BERT-base has 12 encoder layers, 12 self-attention heads, and 768 hidden states ($L=12$, $A=12$, $H=768$). The input of the BERT-base model is a sequence of $M$ tokens, where $M$ is the sequence length. The Input Embedding layer and Positional Encoding layer convert the token sequences into a matrix of dimension $M \times 768$. We choose to use the sequence length of 128, as it is the typical size used by the General Language Understanding Evaluation (GLUE) benchmark~\cite{glue}. The Embedding layer consumes a large amount of memory and much less computation compared to the encoders, so we assume this step is done by the FPGAs that are dedicated to receiving inputs from users and performing data pre-processing.
In this work, we focus on building the encoders.

\subsection{I-BERT}\label{CH-Background-bert}

Integer-BERT~\cite{ibert} is a quantization scheme that quantizes the inference phase of RoBERTa ~\cite{roberta}, which has the same architecture as BERT, with integer-only arithmetic. For the matrix-multiplication operations, I-BERT uses INT8 format for the input matrices and generates an INT32 format output matrix. The INT32 outputs are fed into the Quantization (Quant) module that transforms the INT32 format data into INT8 format data. Based on the results from the GLUE benchmark, the accuracy of I-BERT stays the same or increases compared to the baseline RoBERTa model. In this work, we focus on accelerating the inference phase of RoBERTa-base with the I-BERT quantization scheme and refer to it as the I-BERT base model, which has the same dimensions as BERT-base. The calculations in the I-BERT encoder mainly consist of Matrix-Multiply, Quant, Softmax, LayerNorm, and GELU.

\section{Related Works}\label{CH-RELATED-WORK}

First, we present works on accelerating MPI on FPGAs and compare the difference between previous works and our work. Second, we present works on automation toolchains for deploying machine learning models on multi-FPGAs. Third, we present works on accelerating transformers on FPGAs.

\subsection{MPI Acceleration on FPGAs}
There have been several works to accelerate various MPI functions on FPGAs.

TMD-MPI~\cite{TMDMPI} is the first implementation of the MPI protocol on a heterogeneous platform that includes CPUs and FPGAs. TMD-MPI supports both point-to-point communication and collective communications for communications between CPUs and FPGAs, as well as communications between FPGAs. The TMD-MPE (Message Passing Engine) is a hardware block attached to each Processing  Element (PE) that provides hardware support for TMD-MPI functionalities. One problem with TMD-MPI is the functionalities of TMD-MPI are integrated into the whole platform, and each PE requires one TMD-MPE to support MPI communications, which can increase resource utilization. Another problem of TMD-MPI is that it only supports one communicator, which limits the scalability of the MPI applications.

ZRLMPI~\cite{zrlmpi} is designed to provide a unified programming model for heterogeneous clusters that include CPUs and FPGAs. Instead of letting users write customized HLS codes, ZRLMPI aims to take the existing MPI-based applications and modify the code automatically so that the modified program can run on a heterogeneous cluster. The compiler, called ZRLMPIcc, is used to identify the parts of a program that will run on the FPGAs and convert them to HLS code.

ACCL~\cite{accl} is an open-source library for FPGA-accelerated collective communications. It implements MPI-like functionalities including Send, Receive, Broadcast, Scatter, (All) Gather, (All) Reduce, etc. It functions as an offload engine for host servers. The control plane utilizes the MicroBlaze microprocessor~\cite{microblaze} for handling complex control decisions. However, the use of the MicroBlaze adds extra latency compared to pure hardware control logic. ACCL supports communicators and treats individual FPGAs as a rank and groups FPGAs together as communicators. The protocol adds a 64-byte header to each packet. One problem with ACCL is that the header adds a significant amount of overhead for messages with sizes less than 1MB, and it offsets the benefits of adding hardware support and cannot compete with software MPI. 

Previous works on accelerating MPI using FPGAs have tried to remain compatible with existing codes, act as accelerators for the MPI functions, or add MPI functionalities into switches. Past experience has shown that it is extremely hard to design MPI-compatible hardware on FPGAs. In our case, we want to provide MPI-like functionalities to the communications amongst kernels in a Galapagos cluster, as well as communications between Galapagos Clusters, but we are not trying to be code compatible with any legacy code. Instead of designing MPI-compatible protocols, we design the Galapagos Messaging Interface (GMI) as a collective communication protocol that is optimized for inter-FPGA communications in Galapagos clusters.
We describe GMI in Section~\ref{CH-GMI}.

\subsection{Multi-FPGA Automation Toolchains}\label{CH-Toolchains}

AIgean~\cite{aigean} provides a deployment framework for deploying hls4ml~\cite{hls4ml} machine learning models on FPGA clusters using the Galapagos framework. Hls4ml first takes a trained model and builds hardware HLS cores. AIgean then builds hls4ml-to-Galapagos bridges that convert the HLS arrays to AXI-Stream interfaces. Once the layers are built by hls4ml, AIgean uses a greedy search-based partitioner to place the hls4ml layers into multiple FPGA nodes and wrap the layers into Galapagos kernels by inserting the hls4ml-to-Galapagos bridges between hls4ml layers. Once the kernels are built, AIgean builds the Galapagos cluster using the Galapagos tool flow. 

Similar to the AIgean automation flow, the Cluster Builder we describe in this paper also works as a front-end of the Galapagos tool flow and generates HLS kernels that can be integrated into the Galapagos framework. The main difference between AIgean and the Cluster Builder is that the Cluster Builder can build multiple Galapagos clusters for an application and can provide collective communication capabilities.

\subsection{Transformer Acceleration on FPGAs}

FTRANS~\cite{ftrans} implements single-FPGA transformers based on the block-circulant matrix (BCM) compression techniques. The overall architecture consists of a CPU host and an FPGA accelerator. The host CPU transfers input data to the FPGA for inference through PCIe. For smaller transformers, like shallow transformers, the encoder and decoder layers are implemented as encoder and decoder stacks in a fully unrolled fashion. For larger transformer models like BERT, the encoder and decoder stacks need to be reused. The Processing Element (PE) design is different than conventional matrix-multiplication modules, as it uses an FFT\slash IFFT-based architecture for BCM-based  matrix-multiplication. The use of compression algorithms in the transformer inference phase causes a relatively small accuracy drop for small transformers, but for larger models like BERT, the accuracy drop is about 4.3 percent. 

NPE~\cite{npe} is an FPGA-based overlay accelerator that can be used to accelerate different NLP models. Unlike FTRANS, which unrolls the transformer encoder and decoder layers and implements them on FPGA directly, NPE provides matrix-multiplication modules and nonlinear function modules on top of the FPGA fabric and uses them to accelerate the NLP models in a layer-by-layer fashion. NPE focuses on the BERT-base model as their accelerated model and focuses mainly on accelerating the encoder layers as well. NPE achieves a low inference latency, however, because it reuses the overlay accelerator for each layer, the throughput is relatively low compared with our work.

DFX~\cite{dfx} is a multi-FPGA acceleration system that accelerates the inference phase of GPT-2~\cite{gpt2}. The FPGAs are connected to the host server as offload engines through PCIe. The FPGAs are connected together using a ring topology, and in their experiment, a total of four FPGAs are used in their FPGA cluster. The DFX adopts the intra-layer parallelism method and partitions each layer in the GPT-2~\cite{gpt2} model into the four FPGAs, so that each FPGA runs a portion of a layer and the results will be gathered through the ring network as the synchronization step after each layer. One problem with DFX is that the framework is not flexible, as the FPGAs can only form a ring topology and can only function as offload engines. Moreover, FPGAs in different clusters cannot directly communicate with each other and have to rely on the CPU servers for inter-cluster communication, which adds a high latency overhead.

A recent survey~\cite{kachris2024survey} on hardware accelerators for LLMs also includes the above and several other FPGA-based works.
Other than DFX~\cite{dfx}, none of the prior works presented address the use of multiple FPGAs.
DFX can launch parallel tasks across multiple FPGAs used as offload accelerators, so the FPGAs cannot directly interact, which is a different model than the one used in our work.

FlightLLM~\cite{flightllm:fpga2024} enables efficient processing of compressed LLMs leveraging the DSPs and memory hierarchy of FPGAs.
The spacial sequential architecture (SSR) approach of Zhuang et al.~\cite{zhou:fpga2024} shows how combining sequential and spacial approaches can achieve better latency and throughput on an FPGA.

The most interesting effort related to the work presented in this paper is by Cheng et al.~\cite{zhang:arxiv2023} that does an analytical study of the potential of model-specific spatial acceleration for LLM inference on FPGAs.
The analysis fits very well with the implementation model used by Galapagos where there there are separate streaming kernels for each function being computed.
The availability of an analytical model will be useful when it comes time to develop an efficient transformer implementation that could be deployed using our proposed infrastructure.

Previous works on accelerating Transformers using FPGAs either remain in the single-FPGA domain or utilize multiple FPGAs in a less flexible style. 
The focus of all the previous works is on achieving better performance through algorithms and computational architecture.
In future, all of the previous works could be used as the basis for implementing more efficient multi-FPGA transformers than our proof-of-concept demonstration.

\subsection{Cloud-Scale Machine Learning on FPGAs}
The most well-known implementation of machine learning at cloud scale is Microsoft's Project Brainwave~\cite{brainwave:micro2018}.
Brainwave is based on a soft neural processor implemented in the FPGA and can implement DNN models as hardware microservices with low latency.
The use of the soft processor enables faster implementation since compilation requires instruction generation rather than high-level synthesis.
One of the key strategies of Brainwave is to keep weights in FPGA memory to leverage the high on-chip memory bandwidth.
If there is not enough memory, then another FPGA is used rather than using reduced models or off-chip memory.
This can work because of the low-latency network connectivity.
There has been no further reports of whether the Brainwave approach has been used to implement LLMs.

\section{Scaling Up with Clusters of Clusters}\label{CH-SCALING}

The original Galapagos only supports 256 kernels~\cite{galapagos} in a cluster.
At the time of designing Galapagos, large-scale machine learning models, such as LLMs, were not yet mainstream.
We felt that 256 kernels would be adequate for what we thought would be large applications and 256 was a convenient number in terms of resources.
The main resource consideration is the size of the routing table needed, alongside data retrieval speeds, to specify the routing information of every kernel, as well as the number of bits in the address field of a packet.
Now that we have seen the number of devices used for LLMs, it is clear that much larger numbers of kernels are required and we need a way to scale Galapagos.
In this section, we describe our approach to scaling.

The implementation of scaling ties in closely with the Galapagos Messaging Interface (GMI), to be described in Section~\ref{CH-GMI}, because the addressing of kernels is part of the communications protocol.
The approach we use is a hierarchical addressing scheme much like the idea of subnets in IP addressing~\cite{tannenbaum} where part of the kernel address specifies a particular sub-network and the remaining part of the address specifies the device in the network.
In our case we use a hierarchy that comprises a cluster of Galapagos clusters, i.e., there can be many Galapagos clusters of up to 256 kernels with a layer of communication that handles messages between clusters.
Intra-cluster communication is communication within a single Galapagos cluster.
Inter-cluster communication is communication between different Galapagos clusters.

The original Galapagos Network Layer makes routing decisions based on a routing table with a maximum of 256 IP addresses. However, to support inter-cluster routing, a second table is required for storing additional IP addresses to communicate with kernels in other clusters. 

We make the restriction that for inter-cluster communication, all the incoming messages for each cluster have to arrive at a Gateway kernel (e.g., kernel 0 in each cluster), and direct communications between kernels in different clusters are forbidden. This is due to the consideration of reducing the size of routing tables and saving FPGA on-chip memory. If a kernel in a cluster can send messages to any arbitrary kernel in any other cluster, then for N clusters each having N kernels, a total of $N^2$ IP addresses must be stored in each FPGA. However, if a cluster only sends messages to other clusters through Gateway kernels, then only $2N - 1$ IP addresses will be stored in each FPGA. In this work, we set the maximum number of clusters to 256. Therefore, a total of $256 \times 256 = 65536$ kernels can be accommodated in the enhanced Galapagos framework.

\begin{figure}
    \centerline{
    \includegraphics[width=0.7\linewidth]{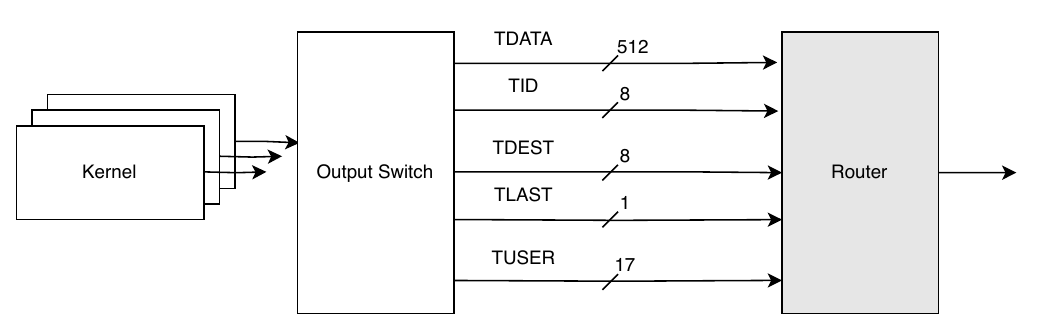}
     }
    \caption{Output Switch and Router in a Galapagos application}
    \label{fig:gmi-router}
\end{figure}

We implement two routing tables using internal BRAMs. The first table is used to access IP addresses for kernels within the same cluster, and the second table is used to access the IP addresses of the Gateway kernels of other clusters. As shown in Figure~\ref{fig:gmi-router},
the modified router requires one additional bit in the TUSER channel (bit16)
to differentiate whether a message is for other clusters or within the same cluster. If it is 0, then the Router accesses IP addresses using the first table to get the IP address for the destination kernel within the cluster. If it is 1, then it accesses IP addresses using the second table to get the IP address for the Gateway kernel in other clusters. The required modification to the Galapagos framework is restricted to only the Network Layer, and other layers remain unchanged showing the benefit of the layered approach used in the Galapagos stack.

\section{The Galapagos Messaging Interface (GMI)}\label{CH-GMI}

As seen in the software world with the use of MPI, it is necessary to have mechanisms to enable different types of communication across the many parallel threads of computation in a large distributed application.
In this section, we present our implementation of a set of communication primitives called the Galapagos Messaging Interface (GMI).
The goal is not to be code compatible with MPI, but to provide an efficient implementation of communication primitives similar to those used in MPI and demonstrated to be the most effective and useful.
It is also important that the implementation fit within the Galapagos approach and philosophy.

\subsection{GMI Architecture}
\begin{figure}
    \centerline{
    \includegraphics[width=0.25\linewidth]{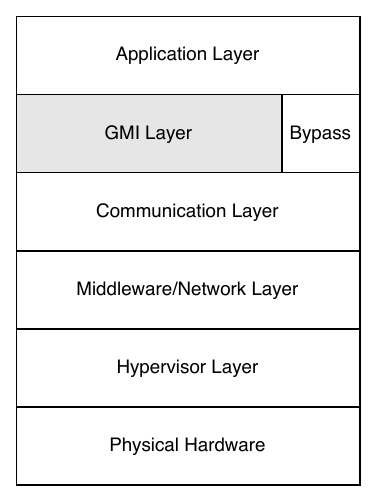}
    }
    \caption{Galapagos stack with the GMI Layer}
    \label{fig:gmi-stack}
\end{figure}

GMI is designed as a communication protocol placed on top of the Communication Layer in the existing Galapagos stack. GMI provides functionalities that are similar to a subset of the MPI protocol and supports various collective communication operations as well as point-to-point communications. Here we define collective communication as the communication that involves a group of kernels in the context of Galapagos Clusters. Figure~\ref{fig:gmi-stack}
shows the logical view of the Galapagos stack with the GMI layer on top of the Communication Layer. The GMI layer handles the communication between the kernels in the Application Layer and the Communication Layer. The GMI layer can also be bypassed when the kernels do not need to use the GMI protocols. This provides flexibility to the applications compared to the previous Communication Layer design, in which all the kernels in the Galapagos cluster follow the same communication protocol, which limits flexibility.

GMI implements functionalities that are similar to MPI but are not compatible with the standard MPI protocol.
The goal is only to add the necessary collective communication support with a minimum hardware cost.
We implement the functionality as GMI kernels, running simultaneously with other kernels in the Application Layer. Since we propose to extend the scalability of the Galapagos framework and allow tens of thousands of kernels to run in a multi-cluster environment, the number of kernels is not a constraint for the design methodology. Thus, it is possible to assign a few kernel IDs to the GMI kernels for handling collective communications. 

In GMI, we implement a basic set of collective communication functions: Broadcast, Reduce, Scatter, and Gather. These basic collective communication functions are sufficient for most scenarios, and other more complex functions like Allreduce or Allgather can be implemented by simply combining the basic functions together. For example, to implement the Allgather operation, first perform a gather operation that gathers the packets from leaf nodes to a root node, then the root node can send the gathered packet to a broadcast module that will broadcast the packet back to all the nodes within the group. 

\begin{figure}
    \centerline{
    \includegraphics[width=0.9\linewidth]{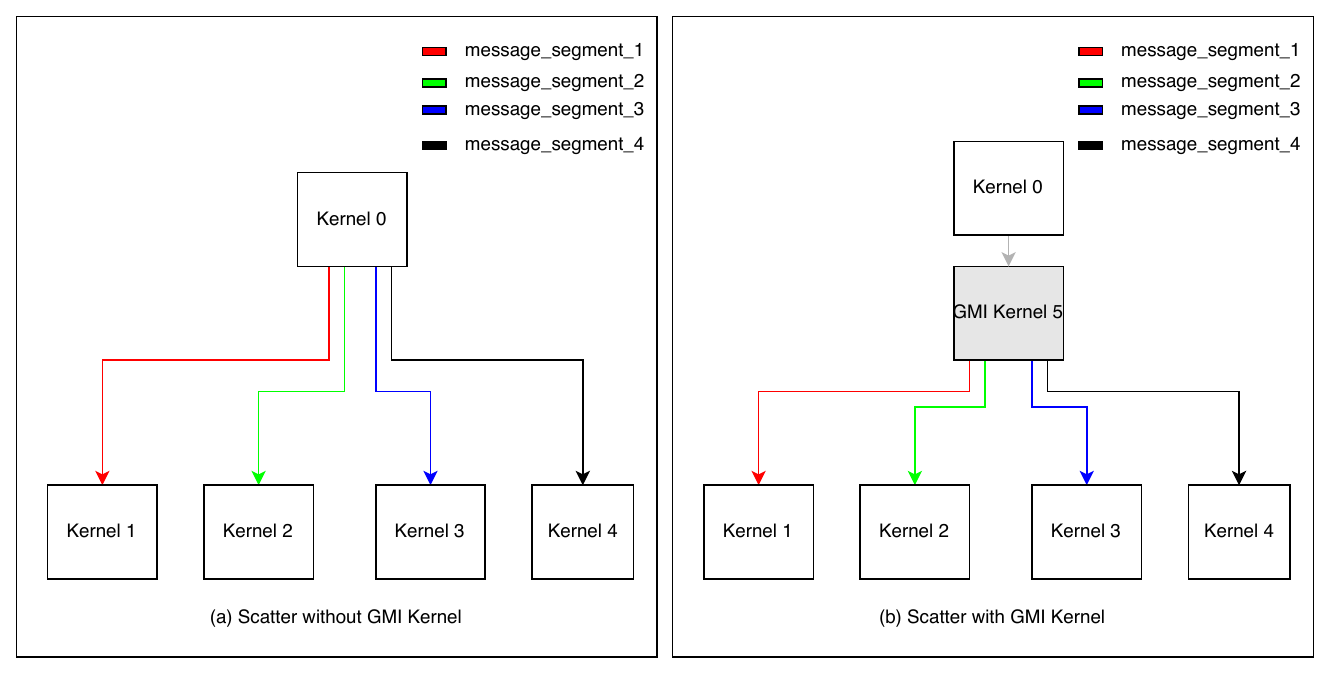}
    }
    \caption{Scatter operation}
    \label{fig:gmi-scatter}
\end{figure}

To free the computation kernels from handling complex communications, the GMI layer provides GMI kernels and inserts them into the multi-kernel graph. For instance, in Figure~\ref{fig:gmi-scatter}(a), Kernel~0 performs the Scatter operation to Kernels 1, 2, 3, and 4. The Scatter functionality has to be implemented in Kernel 0. In Figure~\ref{fig:gmi-scatter}(b), after the insertion of the GMI kernel, Kernel 0 needs only perform computation and then produce the output, and lets the GMI kernel handle the Scatter operation.
This approach decouples the compute function from the communication function.
Figure~\ref{fig:gmi-scatter}(b) also shows how a GMI kernel becomes just another kernel in the connection graph that is input to Galapagos, so there is no modification required for the rest of the Galapagos flow.

Similar to MPI, which has the capability to handle both intra-group and inter-group communication using intra-communicators and inter-communicators, GMI also supports intra-group and inter-group communication.
This addresses the need to extend the 256-kernel limit of Galapagos by being able to implement multiple Galapagos clusters.

In the context of the extended Galapagos framework, a group can be defined as a Galapagos cluster consisting of one or many kernels. 
GMI also provides another level of flexibility for intra-cluster communication and allows the kernels to form subgroups and perform collective operations within subgroups. 
Therefore, within the cluster, there can be several subgroups of kernels performing different collective communications independently. 

The users can either use the Cluster Builder, described in Section~\ref{CH-CLUSTERBUILDER}, to build the GMI applications, or the users can manually configure the GMI kernels by setting the parameters including Kernel ID, Dest, Source, etc. Once the parameters have been configured, the corresponding HLS IP cores can be generated, which can then be used in the Galapagos toolchain.

\subsection{GMI protocol}\label{CH-GMI-protocol}

The GMI protocol is an extremely lightweight protocol that requires only a one-byte header for inter-cluster communication, and no header is required for intra-cluster communication.

For intra-cluster communication, there are two scenarios to consider. For point-to-point communication, there is no need for additional GMI kernels to handle the communication so the GMI Layer can be bypassed.  For collective communication, GMI kernels must be inserted into the multi-kernel graph. Each computation kernel sends its outputs to the desired GMI kernels for performing collective communication. There is no need for an additional header for both point-to-point and collective communication.

For inter-cluster communication, a one-byte header is needed for all the inter-cluster messages. Since a kernel needs to be able to communicate with all kernels in all clusters, it requires the Gateway kernel to have additional information about the destination kernel ID. Thus, the header specifies the destination kernel ID and the kernel can either be a computation kernel or a GMI kernel. The Gateway kernel serves the purpose of forwarding the incoming messages to the correct destination kernel. 

Besides the advantage that there is little overhead for adding the GMI layer in the current Galapagos framework, another benefit of utilizing GMI kernels in multi-kernel applications is the flexibility of placement of the GMI kernels. The GMI kernels can either be placed near the sender side or the receiver side, depending on which provides better bandwidth efficiency. For instance, a Broadcast operation with destination kernels placed on the same FPGA requires less network traffic when the Broadcast GMI kernel is placed on the receiver FPGA, as Broadcast can be done within the receiver FPGA, thus only using the internal bandwidth.

\subsection{Hardware Architecture}\label{CH-GMI-arch}

\begin{figure}
    \centerline{
    \includegraphics[width=0.8\linewidth]{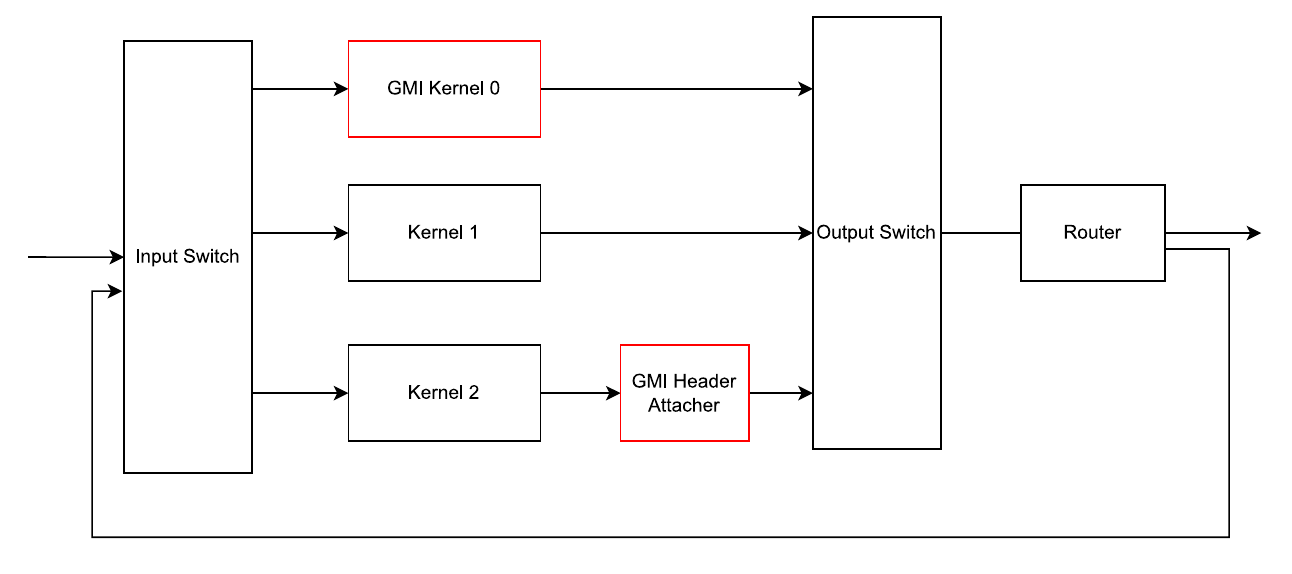}
    }
    \caption{Galapagos application region with GMI Kernel}
    \label{fig:gmi-gp}
\end{figure}

Figure~\ref{fig:gmi-gp} shows an example design of how the GMI kernels are integrated into the Galapagos framework. The GMI kernel is placed in the Application Region the same way as other Galapagos kernels. For intra-cluster communication, there will be no changes to the Galapagos kernels, and for the kernels that send outputs to other clusters, a GMI Header Attacher module is attached to the output stream of the kernel, which adds a one-byte header to the outgoing messages. 

\begin{figure}
    \centerline{
    \includegraphics[width=0.45\linewidth]{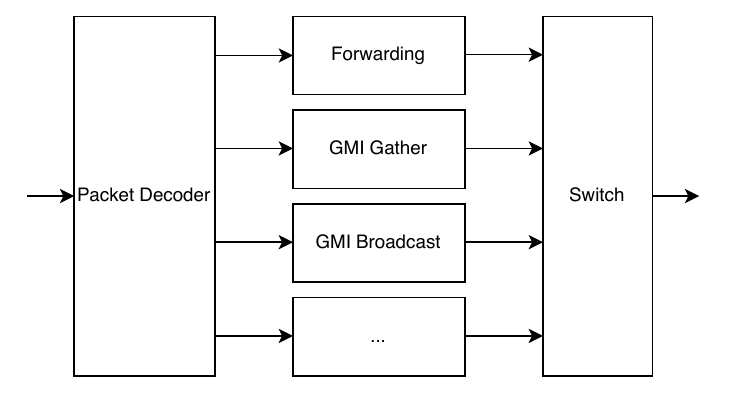}
    }
    \caption{Gateway kernel architecture}
    \label{fig:gmi-gateway}
\end{figure}

To reduce the number of GMI kernels that are used for inter-cluster communications, we integrate the required GMI kernel as part of the Gateway kernel functionalities. The integrated GMI kernels become a sub-module in the Gateway kernel, as shown in Figure~\ref{fig:gmi-gateway}. We define these integrated GMI kernels as virtual kernels, in the sense that they are not physically placed in the Galapagos Application Region. The overall architecture of a Gateway kernel consists of a Packet Decoder, GMI modules for collective operations, a Forwarding module, and an AXI-Stream switch. The Packet Decoder receives the inter-cluster messages, gets the destination kernel ID in the GMI header, then removes the header from the message, and sends the payload to the correct modules. The GMI module performs the collective communication operations and sends the message out through the AXI-Stream switch. The forwarding module handles point-to-point communications.

\section{Cluster Builder}\label{CH-CLUSTERBUILDER}

The Cluster Builder described here aims to provide an automation solution that allows the partitioning of a machine learning model into multiple Galapagos clusters while preserving good performance. 
Other application spaces may require a different clustering approach and a different cluster builder.

Utilizing a multi-cluster approach has an important benefit because it allows the deployment of large-scale models to be managed more easily in datacenters due to the natural partitioning of the application. With the multi-cluster approach, since the inputs to each cluster go to the gateway kernel, we only need one input buffer for each cluster. When one FPGA fails in a cluster, only the cluster that holds the failed FPGA needs to be re-configured, without affecting the rest of the clusters, as the packets that are sent to this cluster will be buffered in the cluster input buffer. This provides one extra layer of flexibility and efficiency for the management of large-scale applications.

In this section we present the Cluster Builder flow we have used, which assumes a fully unrolled machine learning network.
It is the simplest and easiest to implement.
Other strategies for clustering can be explored in future, such as trying to reduce hardware usage by reprogramming parameters when common sub-graphs of hardware are identified that can be scheduled for use at different times.

\begin{figure}[htbp]
    \centerline{
    \includegraphics[width=0.43\linewidth]{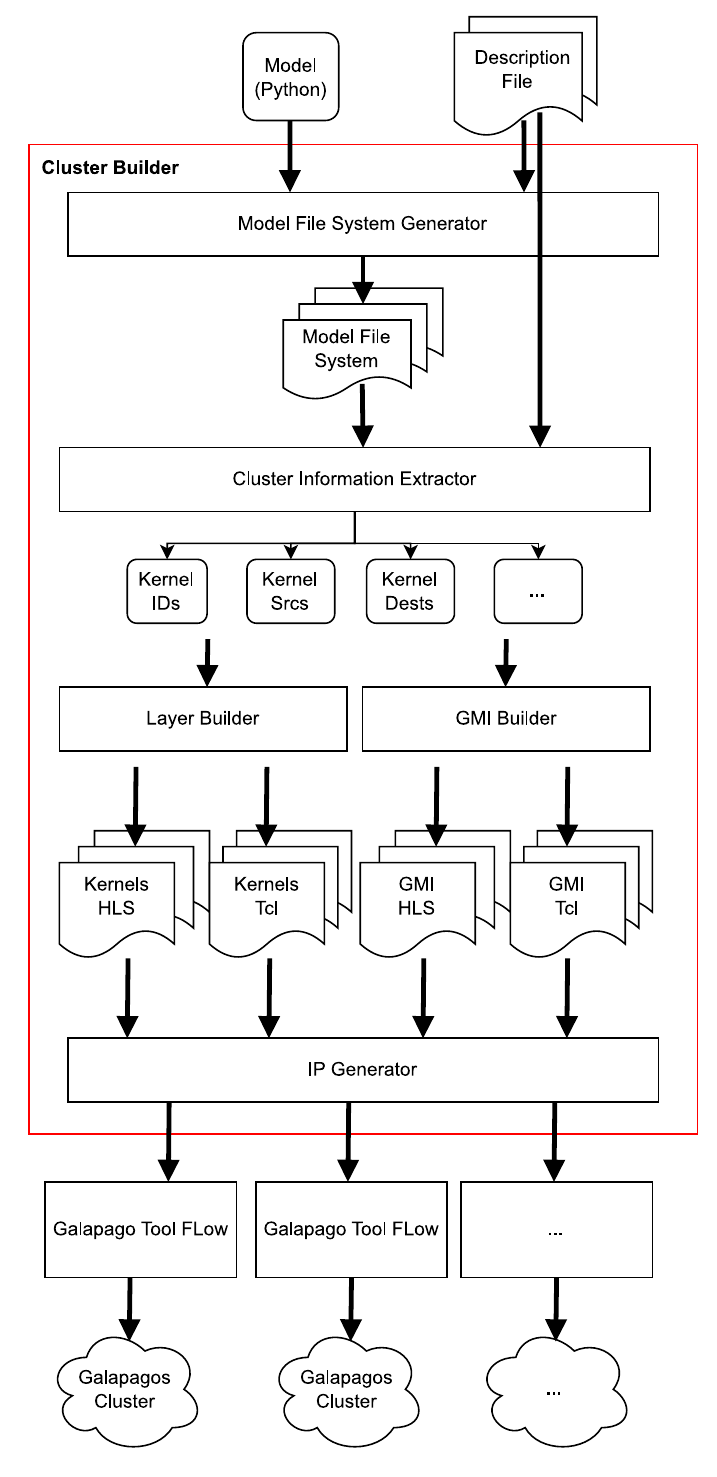}
    }
    \caption{Cluster Builder software architecture}
    \label{fig:cb-arch}
\end{figure}

\subsection{Cluster Builder Flow}\label{CH-CLUSTERBUILDER-arch}
Figure~\ref{fig:cb-arch} shows how the Cluster Builder takes a trained Hugging Face~\cite{huggingface} model and the description files for model layers and clusters as the inputs and generates the HLS code as well as Tcl automation scripts for the Galapagos Tool Flow to build Galapagos clusters. The Cluster Builder consists of several modules including the Model File System Generator, the Cluster Information Extractor, the Layer Builder, the GMI Builder, and the IP Generator. The model itself has several PyTorch modules, such as Linear, LayerNorm, Softmax, etc. We extract the parameters of these modules and store them in a local file system using the Model File System Generator.

We choose the I-BERT base model~\cite{ibert} as the input model to the Cluster Builder. The other input to the Cluster Builder is the Description file, which consists of two JSON files: the Cluster Description File and the Layer Description File. The Cluster Description File specifies the number of clusters and how to partition layers into different clusters. The Layer Description File specifies the connections between layers and the hardware modules to be implemented. It also specifies the configurations and parallelization of the hardware module and allows the user to decide how much FPGA resources to use for it. The Model File System Generator provides a file system that stores the parameters and weights for each module. Once the Model File System Generator creates the file system, the Cluster Information Extractor generates information about \textit{Kernel IDs}, \textit{Kernel Sources}, \textit{Kernel Destinations}, etc. The \textit{Kernel IDs} are divided into three types: the \textit{Compute Kernel ID}, the \textit{Communication Kernel ID}, and the \textit{Virtual Kernel ID}. For each cluster, the \textit{Compute Kernel IDs} are assigned to the computation kernels that implement machine learning model functionalities. The \textit{Communication Kernel IDs} are assigned to the GMI kernels that are inserted between layers within a cluster. The \textit{Virtual Kernel IDs} are assigned to the virtual kernels implemented in the Gateway Kernels. In a cluster, the three types of Kernel IDs form a contiguous ID space from 0 to N-1. 

The outputs from the Cluster Information Extractor are sent to two modules: the Layer Builder and the GMI Builder. The \textit{Kernel IDs}, \textit{Kernel Sources}, \textit{Kernel Destinations}, and \textit{Kernel Types} for compute kernels are used by the Layer Builder for generating the top-level HLS wrappers, Vivado~HLS Tcl scripts, and Vivado Tcl scripts. It calls different handlers for handling different types of kernels. The handlers are responsible for generating parameter files and weight files for specific hardware modules. The Layer Builder generates Vivado~HLS Tcl scripts for creating HLS IP cores later using the Vivado~HLS tool flow. The GMI builder follows a similar procedure as the Layer builder and it generates Tcl scripts for GMI kernels and Gateway kernels. The Tcl scripts provide instructions to Vivado~HLS on setting up IP AXI-Stream interfaces, clock, and reset signals. These interfaces and signals will later be used by Galapagos to connect the IPs to the Galapagos Application region.

The IP Generator module uses the Xilinx Vivado~HLS tool flow and Xilinx Vivado tool flow to generate Galapagos-compatible kernels for each Galapagos Cluster. The generated kernels are sent to the Galapagos tool flow to generate Galapagos clusters.

\section{Accelerating Transformers on Galapagos Clusters}\label{CH-TRANSFORMER}

The transformer~\cite{transformer} is a key component in LLMs.
To test the capability of our enhanced Galapagos framework we wanted to implement a transformer that is large enough to require multiple FPGAs, but small enough that we could manage its design.
We chose I-BERT as our test model for the following reasons. First, it is publicly available on Hugging Face~\cite{huggingface}. Second, I-BERT uses the INT8 data format and uses polynomial approximations for nonlinear functions, which are efficient when implemented on FPGAs. This allows us to implement the I-BERT model directly on FPGAs using the same methods as in their software approach, without any accuracy loss compared to the software version. 
Essentially, we have a very good software reference model for validating our hardware.
In this section we present the I-BERT model and how we map it to hardware using the enhanced Galapagos framework we have developed.
We briefly describe the implementation of some of the hardware layers to help understand some of the results we present and the architecture of the system we have built.
Better optimized implementations are possible but our goal here is to achieve reasonable functionality to demonstrate the capability of our platform for implementation and deployment.

\subsection{Implementing I-BERT}

\begin{figure}
    \centerline{
    \includegraphics[width=1\linewidth]{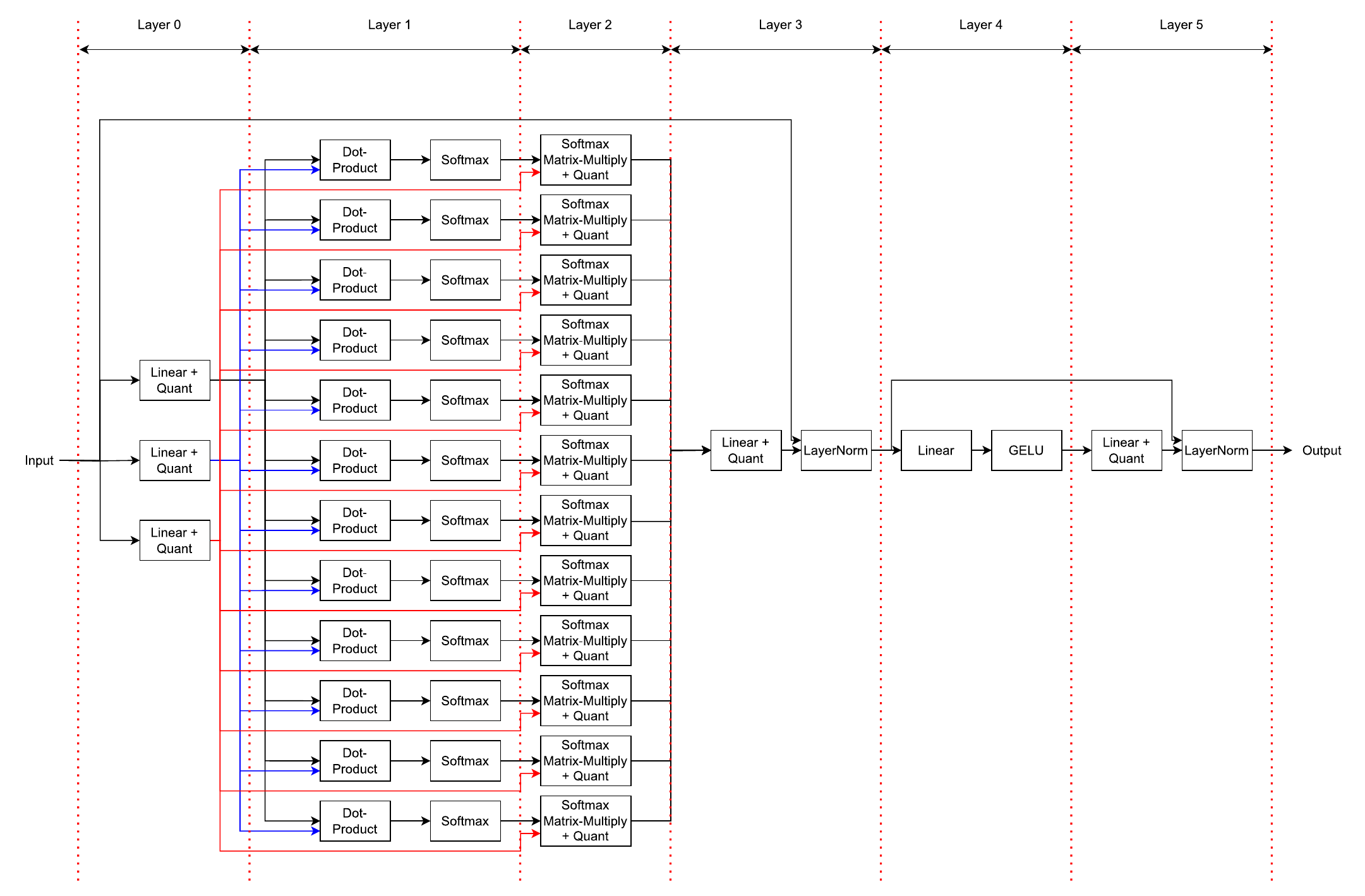}
    }
    \caption{Architecture of the I-BERT encoder }
    \label{fig:ibert-arch}
\end{figure}

Figure~\ref{fig:ibert-arch} shows the architecture of one encoder in the I-BERT model. It consists of six layers, with each layer composed of one or several modules. 
Observe how the communication patterns can be satisfied with the various communicators implemented in the GMI described in Section~\ref{CH-GMI}.

One innovation of our hardware design as a simple attempt to optimize performance is that it does not require any input padding, i.e., short sequences do not need to be padded to the maximum sequence length that can be handled by the hardware.
This reduces the latency when the input sequence length is less than the maximum sequence length. For example, the MRPC micro-benchmark in the GLUE benchmark~\cite{glue} for text classification has an average sequence length of 54, which is much shorter than the maximum sequence length of 128, so not needing to apply padding will reduce the latency of running that benchmark.

To avoid padding, modules that perform matrix-multiply operations need to be designed in different styles. The Linear module in Layer~0 performs matrix-multiply on the input matrix in a row-wise style, and there is no padding required because the input padding simply adds rows to the input matrix. The Dot-Product module in Layer~1 performs dot-product operations on the two input matrices, and only the second input matrix needs minimum padding in the row dimension so that each PE has a matrix column to perform the dot-product operation. This padding can be done immediately after receiving the input matrices. The Dot-Product module also removes the minimum padding for the output matrix, so that the next layers can perform operations on the minimum data required. The Softmax Matrix-Multiply module in Layer~3 receives two input matrices. The first input matrix received from the Softmax module requires minimum padding on the column dimension so that each PE has a matrix row to perform the matrix-multiply operation. The Softmax Matrix-Multiply module also removes the minimum padding for the output matrix. 

We now show the implementation details of the Linear, the Attention Dot-Product, and the Softmax Matrix Multiply module. This is to show how we do the padding and give some understanding for the results we will present.
The implementation details of the Quant, Softmax, LayerNorm, and GELU modules are not shown, as they are implemented the same way as the software version of I-BERT.

\subsubsection{Linear}

\begin{figure}
    \centerline{
    \includegraphics[width=0.5\linewidth]{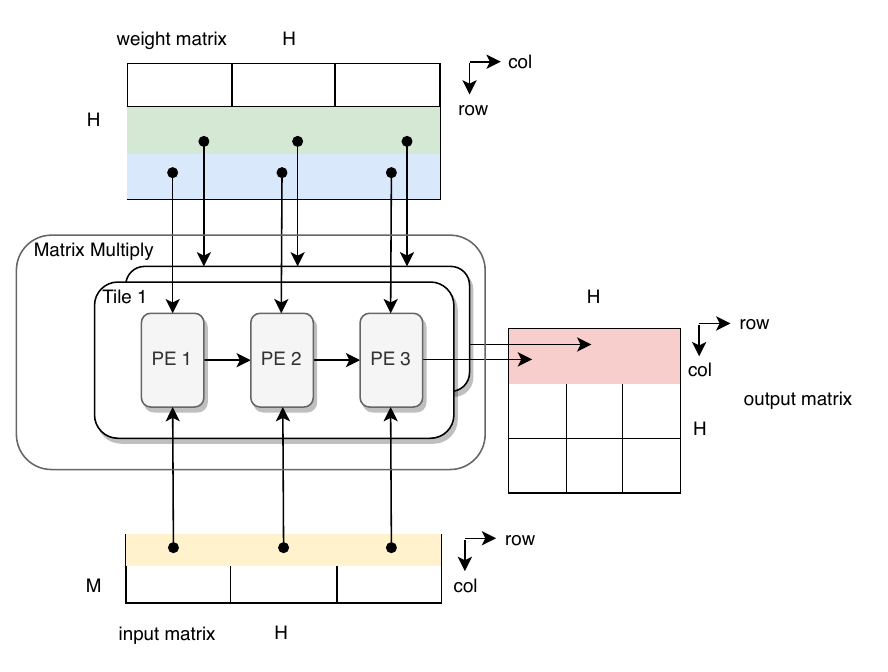}
    }
    \caption{An example of Matrix-Multiply module with two tiles}
    \label{fig:matmul-tiles}
\end{figure}

The Linear module consists of a Matrix Multiply module and a Bias Addition module connected in series in an HLS dataflow region. The Bias Addition module simply adds the bias to the output matrix from the Matrix Multiply module. The input matrix has the dimension of $M \cdot H$, where M is the sentence length and H is the hidden size, and the weight matrix has the dimension of $H \cdot H$. The output matrix has the dimension of $M \cdot H$. Both the input matrix and weight matrix use the INT8 format, while the output matrix uses the INT32 format to avoid overflow. 

The Matrix Multiply module adopts a two-level hierarchical design approach and allows it to be partitioned across multiple kernels. As shown in Figure~\ref{fig:matmul-tiles}, each Matrix Multiply module consists of several Tiles, and each Tile consists of several PEs. The PEs perform partial dot-product on the partial row of the input matrix and the partial column on the weight matrix. The input matrix is streamed in through the AXI-Stream interface and the weight matrix is stored in on-chip memories. The first Tile takes the first column of the weight matrix and the first row of the input matrix, and produces the first data of the output matrix. Similarly, the second Tile takes the second column of the weight matrix and produces the second data of the output matrix. 

There are a few ways of partitioning the Matrix Multiply module into multiple kernels, which can then be placed into multiple FPGAs to utilize more DSP resources. One way is to partition the weight matrix in the column dimension and produce several sub-matrices. Each kernel takes one sub-matrix for computation, and the input matrix is broadcast to the kernels. The output matrix of each kernel can be concatenated together to form the full output matrix. Another way is to partition the row dimension and the output matrix can be accumulated to form the  full output matrix.

\subsubsection{Attention Dot-Product}

The Attention Dot-Product module performs the dot-product operation for an attention head in the attention layer of the encoder. The two input matrices are streamed in through the AXI-Stream interfaces, and have the dimension of $M \cdot K$, where $M$ is the sentence length, $K$ equals $H \mathbin{/} A$, in which $H$ is the hidden size, and $A$ is the number of heads.

\begin{figure}
    \centerline{
    \includegraphics[width=0.8\linewidth]{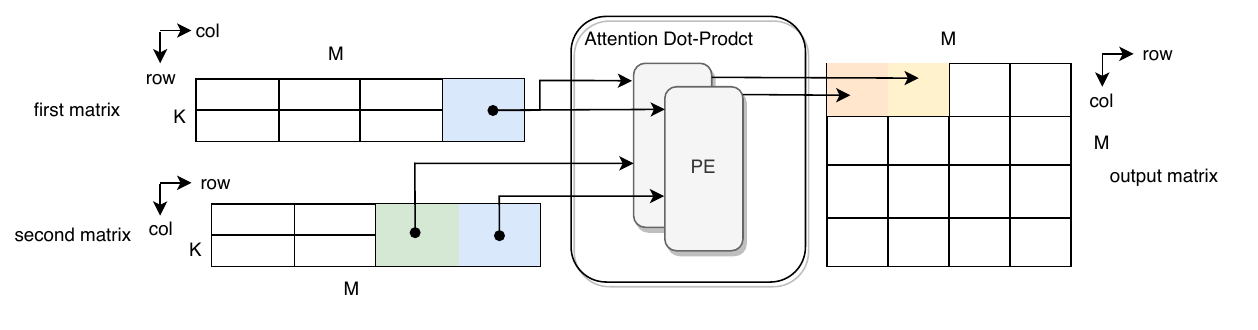}
    }
    \caption{An example of Attention Dot-Product with two PEs}
    \label{fig:dot-product-pe}
\end{figure}

The PEs perform the dot-product on the rows of the first matrix and columns on the second matrix. As shown in Figure~\ref{fig:dot-product-pe},
the row of the first matrix is broadcast to all the PEs, and the columns of the second matrix are scattered to the PEs so that each PE takes a different column. Since each PE needs one column of the second matrix to perform dot-product, the M dimension of the second matrix needs to be padded. The minimum padding is:
\[\textit{NUM}\_\textit{PE} \cdot \lceil M \mathbin{/} \textit{NUM}\_\textit{PE} \rceil\]
where \textit{NUM}\_\textit{PE} is the number of PEs. This padding can be done dynamically when the model is running, and only costs one clock cycle for padding one column.

\subsubsection{Softmax Matrix Multiply}

\begin{figure}
    \centerline{
    \includegraphics[width=0.5\linewidth]{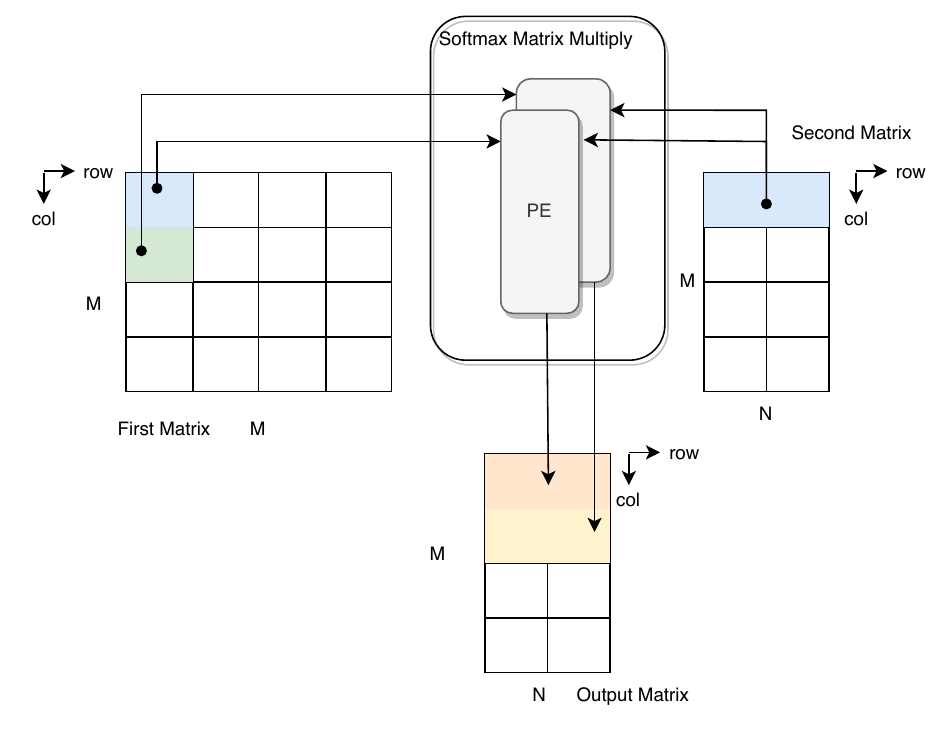}
    }
    \caption{An example of Softmax Matrix Multiply with two PEs}
    \label{fig:softmatmul-pe}
\end{figure}

As shown in Figure~\ref{fig:ibert-arch}, the Softmax Matrix Multiply module performs matrix multiply on the output matrix from the Softmax module from Layer~2 and the Linear module from Layer~0. The matrix from the Softmax module has a dimension of $M \cdot M$, and the matrix from the Linear module has a dimension of $M \cdot N$, where $N$ equals $H \mathbin{/} A$. The output matrix has a dimension of $M \cdot N$. 

Figure~\ref{fig:softmatmul-pe} shows how two PEs perform the matrix multiply. Each PE takes one piece of data from a different row of the first matrix and takes the same row of the second matrix, and produces the partial result of the entire row of the output matrix. One benefit of this calculation style is that the row dimension of the first matrix and the column dimension of the second matrix can be an arbitrary size, which is perfect for the transformer model. For instance, for a sequence of eight tokens, the Softmax Matrix Multiplication module does not require the sequence to be padded to the maximum length, and the PE can iterate only eight times to produce an entire row of the output matrix. Since each PE needs one row of the first matrix, the column dimension of the first matrix needs to be padded. The minimum padding is:
\[\textit{NUM}\_\textit{PE} \cdot \lceil M \mathbin{/} \textit{NUM}\_\textit{PE} \rceil\]
This approach suits well for the second matrix that has a small N dimension, thus each PE can multiply the entire N dimension with data in the first matrix. When N becomes large, the second matrix can be partitioned on the N dimension into smaller matrices, and the Cluster Builder can generate multiple Softmax Matrix Multiply modules that can be placed on multiple FPGAs.

\subsection{Build I-BERT with the Cluster Builder}

\begin{figure}
    \centerline{
    \includegraphics[width=1\linewidth]{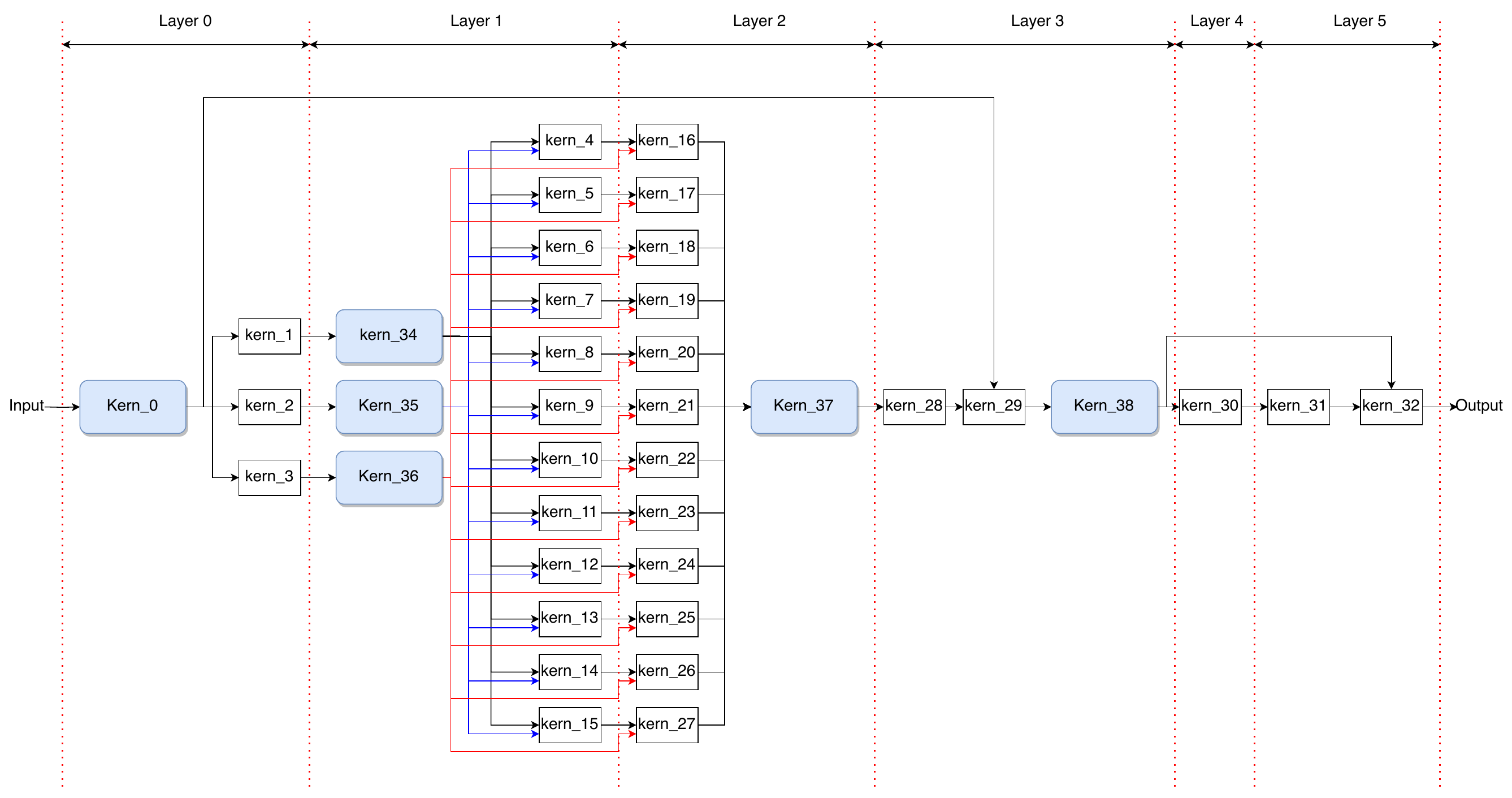}
    }
    \caption{Architecture of the I-BERT encoder built by the Cluster Builder}
    \label{fig:cb-encoder-arch}
\end{figure}

We utilize the Cluster Builder and implement each encoder into one Galapagos cluster.
Figure~\ref{fig:cb-encoder-arch} shows the hardware implementation of an encoder built by the Cluster Builder based on the I-BERT encoder architecture shown in Fig.~\ref{fig:ibert-arch}.
In particular, note the insertion of the GMI kernels.  Kern\_0 implements the Gateway kernel and performs Broadcast. Kern\_1, 2, 3 implements  Linear and Quant. Kern\_4 to 15 implements Dot-Product and Softmax. Kern\_16 to 27 implements Softmax Matrix Multiply and Quant.  Kern\_28 implements Linear and Quant. Kern\_29 implements LayerNorm. Kern\_30 implements Linear and GELU. Kern\_31 implements Linear and Quant. Kern\_32 implements LayerNorm. Kern\_34, 35, and 36 implement Scatter. Kern\_37 implements Gather. Kern\_38 implements Broadcast.

\section{Results from our I-BERT Implementation}\label{CH-RESULT}

In this section, we present results derived from our proof-of-concept system.
We show that the extended Galapagos framework is efficient for accelerating large applications and use the I-BERT base model as our test cast. We present details of the evaluation results for the I-BERT Encoder and layers within each Encoder. We give performance results in terms of resource utilization, latency, and throughput. Using the performance results for one encoder, we develop a latency model derived from the observed behavior of our proof-of-concept implementation, which will be used in Section~\ref{CH-VERSAL} when we do a latency estimate using the recent Versal FPGA.
We make some comparisons with some modern GPUs but with this proof-of-concept these performance comparisons are only provided to give some perspective on the performance being achieved, keeping in mind that the FPGAs in our testbed are not state-of-the-art.

\subsection{Hardware Testbed and Software Tools}

We use up to seven Xilinx Sidewinder-100~\cite{sidewinder} boards, which use an XCZU19EG UltraScale+ FPGA~\cite{zu19eg}, as our test platform for the experiments. All the Sidewinder boards are connected to a DELL Z9100-ON 100G switch~\cite{dellswitch}. The CPU server, with 24 Intel Xeon E5-2650 v4 CPU cores running at 2.2GHz, is connected to a Dell S4048-ON 10G switch, and the 10G switch is connected to the DELL Z9100-ON 100G switch. We use Vivado 2019.1 and Vivado\_HLS 2019.1 as our tools for generating IP cores and bitstreams.

\subsection{I-BERT Evaluation}

For this evaluation, we use six FPGAs to accelerate one encoder and one extra FPGA during the evaluation to provide inputs and receive outputs for the encoder at 100~Gbps, which emulates how the encoder would be connected in the full encoder chain.
We use the GLUE~\cite{glue} benchmark for testing the functionalities and performance of the multi-FPGA I-BERT model. Since we use the same algorithms as the original software to implement the Quant, Softmax, LayerNorm, and GELU modules, the accuracy of our design remains the same as the software version. We confirmed that our design produces exactly the same output as the software version of I-BERT.

\subsubsection{Resource Utilization}\label{CH-ibert-eval-res}

\begin{figure}
    \centerline{
    \includegraphics[width=0.65\linewidth]{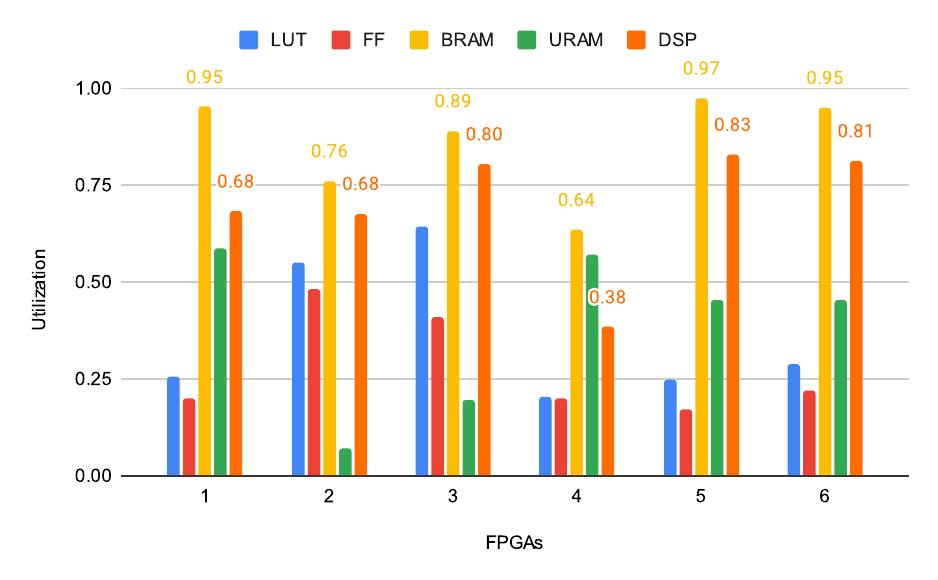}
    }
    \caption{Resource utilization of the FPGAs}
    \label{fig:ibert-res-fpga}
\end{figure}

Figure~\ref{fig:ibert-res-fpga} shows the resource utilization for the six FPGAs. The limiting resource for each FPGA is BRAM. This is because we need to attach AXI-Stream FIFOs to the front and end of each kernel, and to avoid overflow, we need to make each FIFO large enough to hold at least one matrix. For the matrix of dimension $128 \times 768$, we need about 43 18Kb BRAMs to avoid overflow. Moreover, all the weights and parameters in the model are also stored in the on-chip memory. This caused the BRAM resources to be consumed quickly. For FPGAs 1, 5, and 6, the BRAM usage is about 90\%. In terms of DSP usage, FPGAs 3, 5, and 6 use more than 80\% of total DSPs, FPGAs 1 and 2 use about 68\% of total DSPs, and FPGA 4 uses 38\% of DSPs.

\subsubsection{Latency}\label{CH-TRANSFORMER-eval-latency}

\begin{figure}
    \centerline{
    \includegraphics[width=0.8\linewidth]{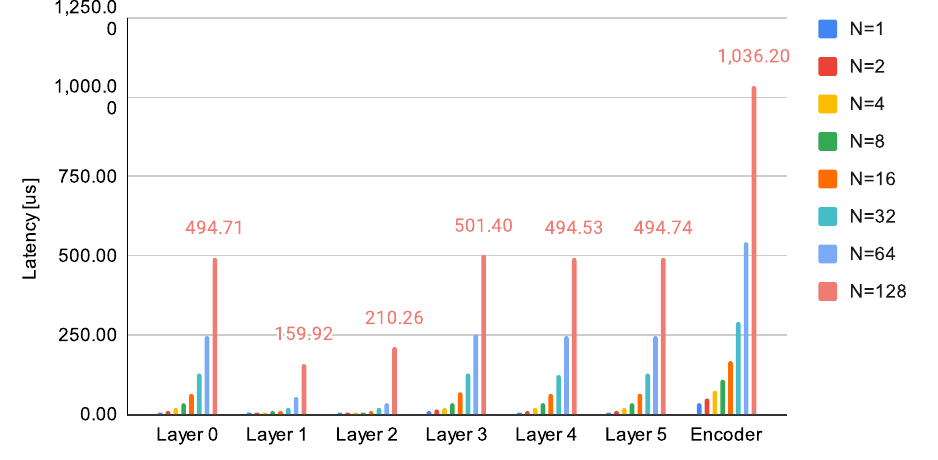}
    }
    \caption{Latency for six-FPGA accelerated I-BERT encoder and layers within the encoder}
    \label{fig:ibert-latency-bar}
\end{figure}

Figure~\ref{fig:ibert-latency-bar} shows the inference latency for the encoder and the six layers in each encoder. Layers 0, 3, 4, and 5 have almost the same latency for sequence lengths from 1 to 128. When the sequence length is 128, the latency of the full encoder is about twice the latency of Layers 0, 3, 4, and 5.

\begin{figure}
    \centerline{
    \includegraphics[width=0.7\linewidth]{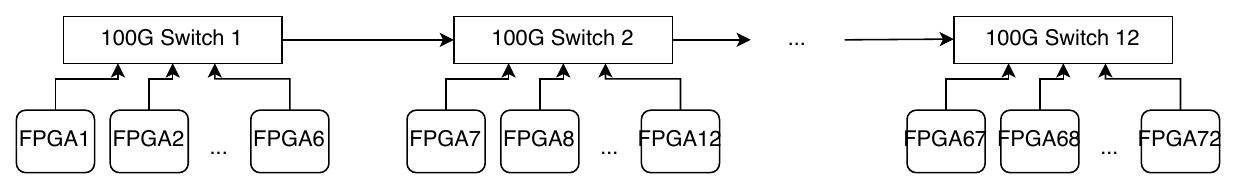}
    }
    \caption{Platform configuration for 72 FPGAs}
    \label{fig:ibert-72fpga}
\end{figure}

Based on the latency of one encoder, we estimate the overall latency of the full I-BERT model. Since each encoder requires six FPGAs, then a total of 72 Sidewinder boards are required to implement all 12 encoder layers. We assume that the Sidewinder boards are connected to a 100G switch with each switch only connecting six Sidewinders for simplicity, as shown in Figure~\ref{fig:ibert-72fpga}. The 12 100G switches are connected serially.

\begin{figure}
    \centerline{
    \includegraphics[width=0.6\linewidth]{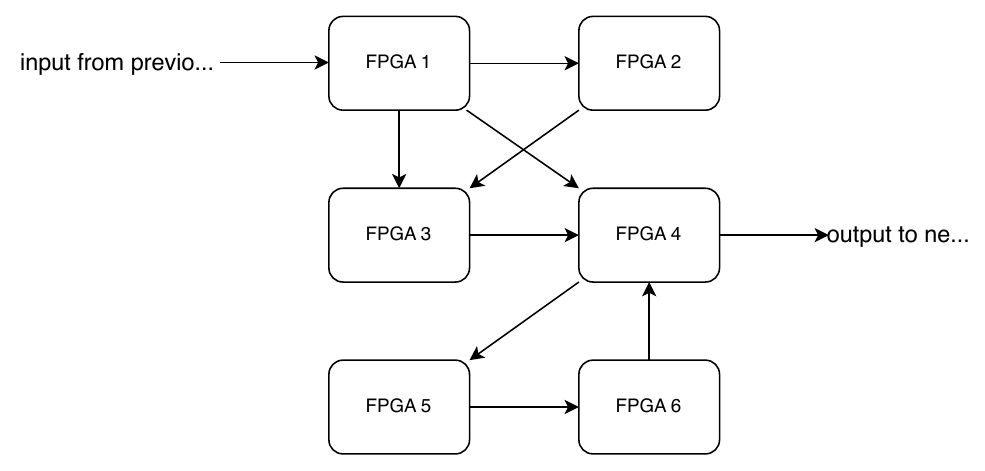}
    }
    \caption{Data transfer pattern between six FPGAs for accelerating one encoder}
    \label{fig:ibert-setup}
\end{figure}

\begin{figure}
    \centerline{
    \includegraphics[width=0.6\linewidth]{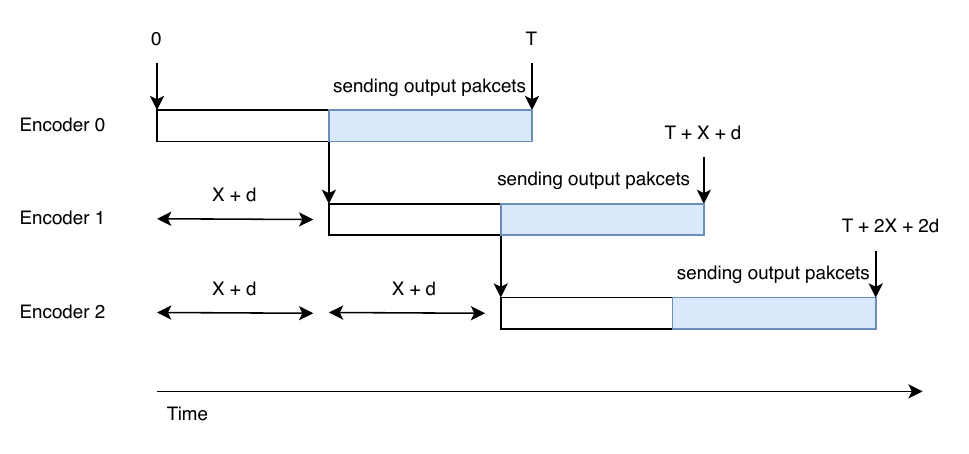}
    }
    \caption{Estimate latency for 72-FPGA accelerated I-BERT}
    \label{fig:ibert-est-latency}
\end{figure}

Figure~\ref{fig:ibert-setup} shows how data is transferred between the six FPGAs for one encoder. FPGA~1 receives the inputs from the previous encoder, whose data generation behavior is simulated by the evaluation FPGA. FPGA~4 sends the outputs to the next encoder and in our experiment, the next encoder is simulated by the evaluation FPGA as well. To estimate the overall latency, we use the equation:
\begin{equation}
\label{eqn:est-latency}
 T + (L - 1) \times (X + d) 
\end{equation}
where $T$ is the inference latency of one encoder, $L$ is the number of encoders, $X$ is the latency when the encoder generates the first data, and $d$ is the latency between the 100G switches. Figure~\ref{fig:ibert-est-latency} shows how the equation is derived. The evaluation FPGA sends the first packet at cycle 0 to Encoder~0, and after $X + d$ cycles, Encoder~1 receives the first packet generated by Encoder 0. Encoder 0 generates the packets with an interval of $I$ cycles. At cycle $T$, Encoder 0 generates the last packet. Since each encoder has the same architecture, we use the same $T$ and $X$ for all encoders.

To simulate the data generation behavior of the previous encoder using the evaluation FPGA, we need to measure the interval at which each encoder generates packets. We first set the evaluation FPGA to send packets at an interval of 12, as each packet contains 12 flits and requires 12 cycles to transfer one packet. This means the evaluation FPGA sends another packet immediately after one packet is sent. We then measure the values for $X$, $T$, and $I$. Then, we set the evaluation FPGA to send packets at an interval of the measured $I$, and measure the values for $X$, $T$, and $I$ again. 
We found the measured values for $X$, $T$, and $I$ stay the same as the first measurement, which suggests the interval of packets does not affect the overall latency. Table~\ref{tab:ibert-cycle} shows the latency measured in clock cycles for $X$, $T$, and $I$. 

We measure the latency between the 100G switches, $d$, to be about 1.1 $\mu s$. Based on the measured values of $X$, $T$, and $d$, the estimated overall latency is shown in Table~\ref{tab:ibert-est-latency}. Based on the fact that not all input sequences have the same sequence length, and the fact that our design does not require any input paddings, we can use the average latency as the estimated latency. We analyzed the input sequences of the GLUE~\cite{glue} benchmark and found that the average sequence length is 38. Based on the average sequence length, we calculated the estimated average latency to be 2.58 ms.

\begin{table}
\centering
\caption{Components of encoder latency, measured in clock cycles}
\label{tab:ibert-cycle}
\begin{tabular}{c|llllllll|}
\cline{2-9}
\multicolumn{1}{l|}{}           & \multicolumn{8}{c|}{Sequence Length}                                                                                                                                                                              \\ \hline
\multicolumn{1}{|l|}{Component} & \multicolumn{1}{l|}{1}    & \multicolumn{1}{l|}{2}     & \multicolumn{1}{l|}{4}     & \multicolumn{1}{l|}{8}     & \multicolumn{1}{l|}{16}    & \multicolumn{1}{l|}{32}    & \multicolumn{1}{l|}{64}     & 128    \\ \hline
\multicolumn{1}{|c|}{X}         & \multicolumn{1}{l|}{6936} & \multicolumn{1}{l|}{10455} & \multicolumn{1}{l|}{13769} & \multicolumn{1}{l|}{17122} & \multicolumn{1}{l|}{23393} & \multicolumn{1}{l|}{35828} & \multicolumn{1}{l|}{61121}  & 111708 \\ \hline
\multicolumn{1}{|c|}{T}         & \multicolumn{1}{l|}{6936} & \multicolumn{1}{l|}{11004} & \multicolumn{1}{l|}{15869} & \multicolumn{1}{l|}{22318} & \multicolumn{1}{l|}{34781} & \multicolumn{1}{l|}{59600} & \multicolumn{1}{l|}{109660} & 209789 \\ \hline
\multicolumn{1}{|c|}{I}         & \multicolumn{1}{l|}{0}    & \multicolumn{1}{l|}{275} & \multicolumn{1}{l|}{525}   & \multicolumn{1}{l|}{650} & \multicolumn{1}{l|}{712} & \multicolumn{1}{l|}{743} & \multicolumn{1}{l|}{759}  & 767  \\ \hline
\end{tabular}
\end{table}

\begin{table}
\centering
\caption{Estimated latency of I-BERT, measured in microseconds}
\label{tab:ibert-est-latency}
\begin{tabular}{l|llllllll|}
\cline{2-9}
                                   & \multicolumn{8}{c|}{Sequence Length}                                                                                                                                                                             \\ \hline
\multicolumn{1}{|l|}{Latency (ms)} & \multicolumn{1}{l|}{1}     & \multicolumn{1}{l|}{2}     & \multicolumn{1}{l|}{4}     & \multicolumn{1}{l|}{8}     & \multicolumn{1}{l|}{16}    & \multicolumn{1}{l|}{32}    & \multicolumn{1}{l|}{64}    & 128   \\ \hline
\multicolumn{1}{|l|}{I-BERT}       & \multicolumn{1}{l|}{0.416} & \multicolumn{1}{l|}{0.630} & \multicolumn{1}{l|}{0.837} & \multicolumn{1}{l|}{1.053} & \multicolumn{1}{l|}{1.461} & \multicolumn{1}{l|}{2.269} & \multicolumn{1}{l|}{3.910} & 7.193 \\ \hline
\end{tabular}
\end{table}

\begin{table}
\centering
\caption{Latency (ms) of our work compared with NVIDIA T4 GPU, NVIDIA A100 GPU, and NPE}
\label{tab:ibert-latency-compare}
\begin{tabular}{|l|r|r|l|l|l|}
\hline
                 & \multicolumn{1}{l|}{NVIDIA T4} & \multicolumn{1}{l|}{NVIDIA A100} & \multicolumn{1}{c|}{NPE (FPGA)} & \multicolumn{1}{c|}{\begin{tabular}[c]{@{}c@{}}Our Design \\ (padding)\end{tabular}} & \multicolumn{1}{c|}{\begin{tabular}[c]{@{}c@{}}Our Design \\ (no padding)\end{tabular}} \\ \hline
Latency (ms)     & 1.66                           & 0.77                             & 13.96                    & 7.19                                                                                 & 2.58                                                                                    \\ \hline
Relative Speedup & 8.4                           & 18.13                            & 1 (baseline)             & 1.94                                                                                 & 5.4                                                                                     \\ \hline
\end{tabular}
\end{table}

We compare the inference latency of our work with BERT acceleration using the NVIDIA T4 GPU~\cite{nv-t4}, the NVIDIA A100 GPU~\cite{nv-a100}, and NPE~\cite{npe} with 8-bit Matrix Multiplies, shown in Table~\ref{tab:ibert-latency-compare}. The latencies for the NVIDIA T4 and A100 are from the NVIDIA TensorRT report~\cite{nv-bert}. All the works are compared using the same BERT base architecture, 8-bit data format, max sequence length of 128, and batch size of 1. NPE~\cite{npe} currently achieves the lowest latency of 13.96 ms among the single FPGA accelerators. When the input sequences are padded to the maximum sequence length, we achieve a relative speedup of 1.94 compared to NPE. When the input sequences are not padded, we achieve a relative speedup of 5.4 compared to NPE. Our proof-of-concept cannot compete with the NVIDIA T4 implementation when inputs are padded, however, when inputs are not padded, our work is more comparable. The NVIDIA A100 achieves a much higher speedup than all the other implementations, and our work is not as comparable. 

The comparison between our work and the other FPGA implementations is not straightforward, as we have more resources because we use more FPGAs. The important point here is that our approach demonstrates the potential for reducing latency when multiple FPGAs are used and that the network interconnection between FPGAs is not a limiting factor.
With these measurements we are able to develop a latency model for multiple FPGAs based on our working proof-of-concept.

\subsubsection{Throughput}

\begin{figure}
    \centerline{
    \includegraphics[width=0.8\linewidth]{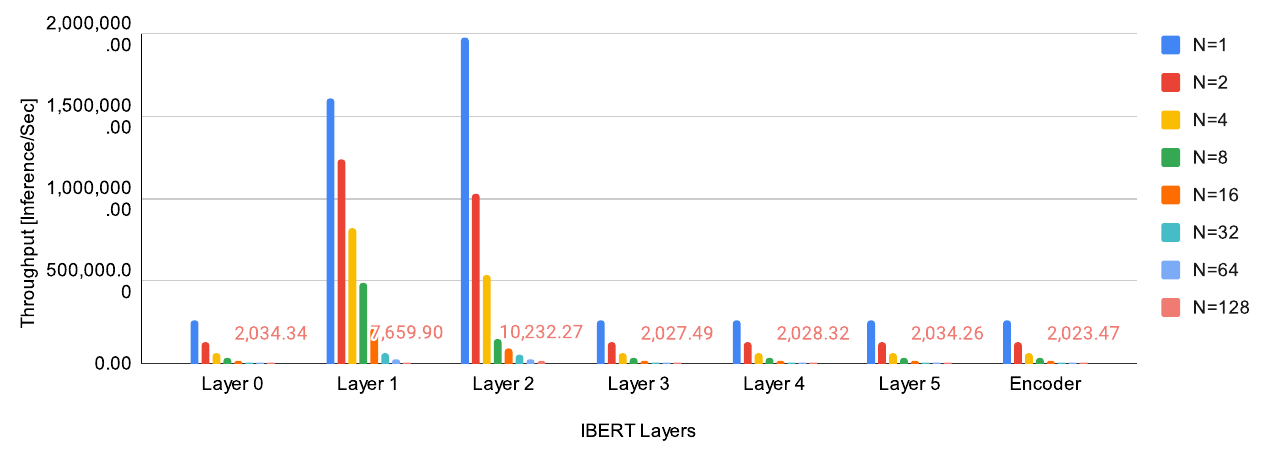}
    }
    \caption{Throughput for Six-FPGA Accelerated I-BERT Encoder and Layers within the Encoder}
    \label{fig:ibert-throughput-bar}
\end{figure}

Figure~\ref{fig:ibert-throughput-bar} shows the throughput results, which are measured in inferences\slash second. Layers 0, 3, 4, 5, and the full encoder have very close throughput, while Layers 1 and 2 have much higher throughput for all sequence lengths. The throughput for the encoder at a sequence length of 128 is 2023.47 inferences\slash second, which is about the same as Layers 0, 3, 4, and 5. This is expected as the overall throughput should be the same as the layers with the lowest throughput. The overall throughput for I-BERT should also be the same as one encoder, as each encoder is connected serially in a dataflow fashion. For the average sequence length of 38, the estimated average throughput is 6802.26 inferences\slash second.

\begin{table}
\centering
\caption{Throughput (inferences\slash second) of our work compared with FTRANS and NPE}
\label{tab:ibert-throughput-compare}
\begin{tabular}{|l|l|l|l|l|}
\hline
                              & \multicolumn{1}{c|}{FTRANS} & \multicolumn{1}{c|}{NPE} & \multicolumn{1}{c|}{\begin{tabular}[c]{@{}c@{}}Our Design \\ (padding)\end{tabular}} & \multicolumn{1}{c|}{\begin{tabular}[c]{@{}c@{}}Our Design \\ (no padding)\end{tabular}} \\ \hline
Throughput (inference/second) & 101.79                      & 135.14                   & 4120.6                                                                               & 6802.26                                                                                 \\ \hline
Relative Speedup              & 0.75                        & 1 (baseline)             & 30.5                                                                                 & 50.3                                                                                    \\ \hline
\end{tabular}
\end{table}

We compare the inference throughput of our work with FTRANS~\cite{ftrans} and NPE~\cite{npe}, shown in Table~\ref{tab:ibert-throughput-compare}. All the works use the max sequence length of 64. NPE achieves the highest throughput of 135.14 inferences\slash second among the single FPGA accelerators. When the input sequences are padded, we achieve a relative speedup of 30.5 compared to NPE. When the input sequences are not padded, we achieve a relative speedup of 50.3 compared to NPE. 
Again, the improvements are achieved by having more FPGA resources available.

\begin{table}
\centering
\caption{Throughput (inferences\slash second) of our work compared with NVIDIA T4 and NVIDIA A100}
\label{tab:ibert-throughput-compare-gpu}
\begin{tabular}{|l|l|l|l|l|}
\hline
                              & \multicolumn{1}{c|}{NVIDIA T4} & \multicolumn{1}{c|}{NVIDIA A100} & \multicolumn{1}{c|}{\begin{tabular}[c]{@{}c@{}}Our Design \\ (padding)\end{tabular}} & \multicolumn{1}{c|}{\begin{tabular}[c]{@{}c@{}}Our Design \\ (no padding)\end{tabular}} \\ \hline
Throughput (inference/second) & 1581.2                         & 11962.6                          & 2,023.47                                                                             & 6802.26                                                                                 \\ \hline
Relative Speedup              & 1 (baseline)                   & 7.56                             & 1.28                                                                                 & 4.3                                                                                     \\ \hline
\end{tabular}
\end{table}

We also compare the inference throughput of our work with the NVIDIA T4 and the NVIDIA A100, shown in Table~\ref{tab:ibert-throughput-compare-gpu}. All the works are compared using the max sequence length of 128. The throughput for the GPUs is derived using the latency for the largest batch size of 128, as the GPUs can achieve much higher throughput at high batch sizes. For instance, the NVIDIA T4 has a latency of 80.95 ms for a batch size of 128, which means the average latency for one input is $80.95 ms \mathbin{/} 128 = 0.632 ms$. The throughput is $1 \mathbin{/} 0.632 ms=1581.2$ inference\slash second. When the input sequences are padded, we achieve a relative speedup of 1.28 compared to the NVIDIA T4. When the input sequences are not padded, we achieve a relative speedup of 4.3 compared to NVIDIA T4. The result shows our work can outperform NVIDIA T4, but is still not comparable to NVIDIA A100, as the A100 GPU has much more computing power.

However, there is a subtle nuance in this throughput comparison.
If latency is also important, then when running with batching, the results for all inputs in a batch are available when the batch completes, which is much longer than the batch-1 latencies of the GPUs.
Our implementation can be considered to be a long pipeline, so outputs are produced at the same rate as the inputs are provided with the batch-1 latencies shown in Table~\ref{tab:ibert-latency-compare}.

It is also important to realize the comparisons with the GPUs at this point only give some indication of where the proof-of-concept platform stands relative to the performance of recent GPUs.
The proof-of-concept platform does not use state-of-the-art FPGAs.
However, the current platform does provide a basis for doing an estimation of performance using more recently introduced FPGAs.
\section{Estimating Transformer Performance on AMD Versal ACAP}\label{CH-VERSAL}

For our main goal of building a platform that is capable of implementing very large applications, it is fine to use a smaller transformer implemented on smaller FPGAs to validate the platform.
The primary goal of our first implementation is to show that we can implement a transformer on multiple FPGAs using the enhanced Galapagos platform and the Cluster Builder tool.
We feel that our approach is successful and given the results we have achieved so far, it is now possible to estimate what it would take to implement I-BERT on the latest FPGA architecture to see how many FPGAs are required and what performance might be achieved using the latest generation of FPGAs.
This will give a better sense of whether the performance using modern FPGAs can be competitive with GPUs.

In this section, we discuss and estimate the inference performance for the I-BERT base model using AMD Versal adaptive compute acceleration platforms (ACAPs)~\cite{versal}. No actual implementation has been done.  The work presented here is still a preliminary study to determine whether an implementation is feasible and estimate how it might perform to help decide whether it is even worth going forward with further implementations and development.

We first give background on the architecture of Versal ACAPs. Then, we give details on the required modifications to the Galapagos framework so that Galapagos can support the Versal architecture. Last, we show details on mapping I-BERT to multiple Versal devices and present the estimated latency for I-BERT.

\subsection{Background}\label{CH-VERSAL-backround and related}

\begin{figure}
    \centerline{
    \includegraphics[width=0.4\linewidth,height=!]{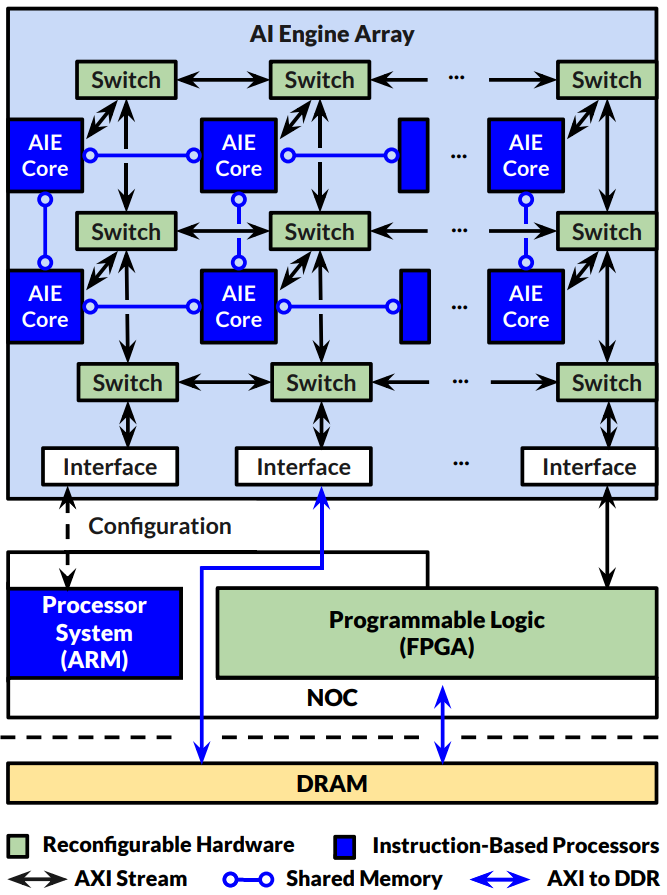}
    }
    \caption{Versal ACAP Architecture, image from~\cite{charm}}
    \label{fig:versal-arch}
\end{figure}

The Versal Adaptive Compute Acceleration Platform (ACAP)~\cite{versal} is an integrated programmable platform that combines Programmable Logic (PL), AI Engines (AIEs), a Processor System (PS), a Network-on-Chip (NoC), and various hardened IPs. Figure~\ref{fig:versal-arch} shows the architecture of the Versal ACAP with a connection to an external DRAM as would be found on a typical board. The VCK190 evaluation board~\cite{vck190} uses a XCVC1902-2MSEVSVA2197 device with 400 AI Engines (AIE) arranged as an $8\times 50$ grid.  Each AIE has a VLIW processor that can operate at a frequency of 1 GHz and  support vector operations up to 1024 bits. The PS consists of ARM processors that can run software programs. The PL consists of DSPs, BRAMs, URAMs, and LUT logic. The NoC connects the PS, PL, and AIE together. There are also 39 interface tiles in the AIE that connect the PL and the AIE directly through PLIOs. On the VCK190, the bandwidth for communication from the PL to the AIEs is about 1.2 TB\slash s, and from AIE to PL is about 0.9 TB\slash s. The DRAM provides a peak bandwidth of 25.6 GB\slash s. Each AIE has 32 KB of data memory and a 2 KB vector register file. The AIE processor is capable of fetching $2\times 256$ bits of data from the data memory in each clock cycle. The AIEs can communicate with each other through AXI-Stream switches, which are capable of sending 32 bits of data per cycle. The AXI-Stream switch provides both circuit-switched connections and packet-switched connections. Circuit-switched connections allow deterministic communications and support broadcast among AIEs. Packet-switched connections require each packet to have a header that specifies the destination and allows the packet to be dynamically routed to different AIEs.  Neighboring AIEs can communicate through their local memories at a higher bandwidth with 256-bit reads and writes.

\subsection{Galapagos for Versal ACAP}\label{CH-VERSAL-gp}

\begin{figure}
    \centerline{
    \includegraphics[width=0.5\linewidth]{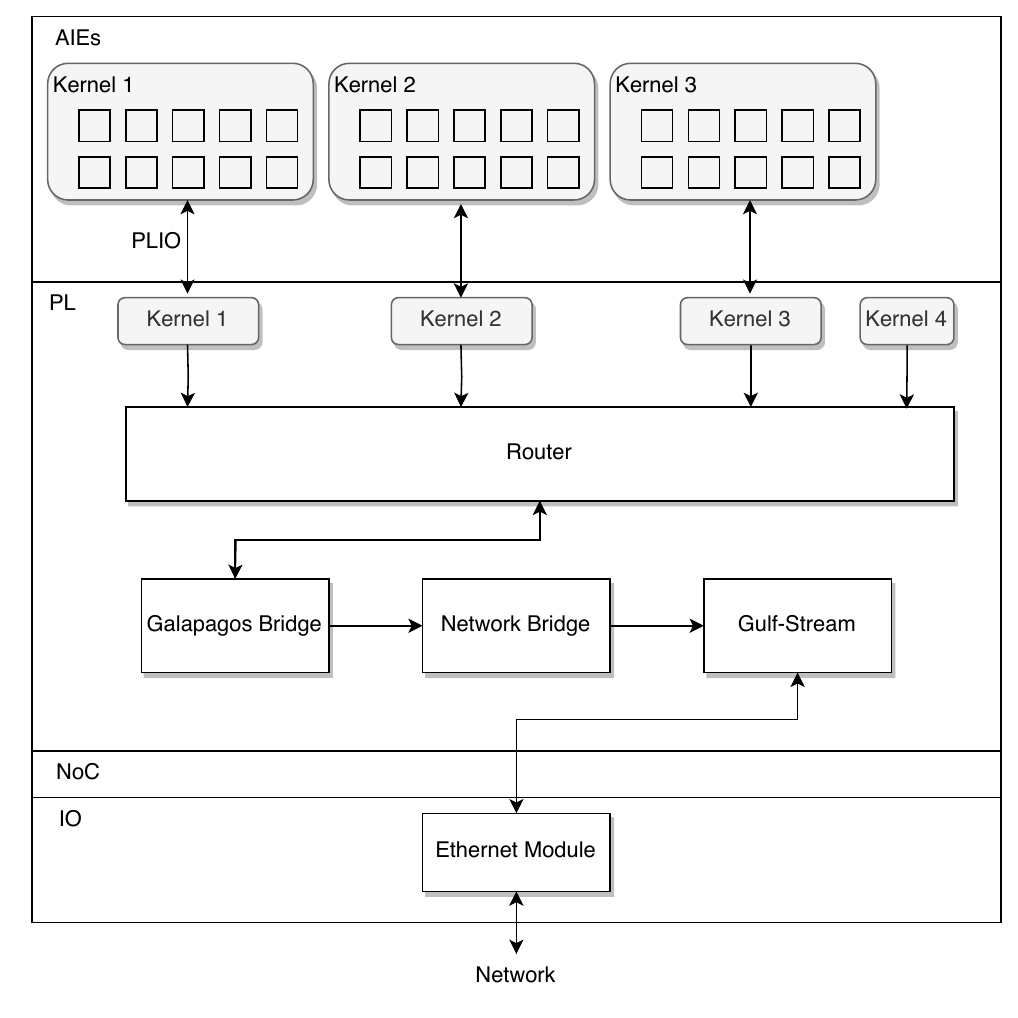}
    }
    \caption{Modified Galapagos Hardware Architecture for Versal Devices}
    \label{fig:versal-gp}
\end{figure}

Since we use Galapagos as our underlying framework, we need a modified Galapagos for supporting Versal cards.
The main goal is to leverage the AIEs when computation is required, such as for matrix multiplication.
Figure~\ref{fig:versal-gp} shows the proposed modified Galapagos hardware architecture that can be implemented on the Versal. For the kernels that use the AIEs, we partition the kernel into two parts, the AIE part, and the PL part. The PL part of the kernel receives Galapagos packets from the Router, in the same way as the current Galapagos framework. To send the packet to the AIE part of the kernel, the PL part converts the Galapagos packet to an AIE-compatible packet. For kernels that do not require AIEs, we can simply write HDL or HLS kernels and connect them to the Router. For instance, in Figure~\ref{fig:versal-gp} Kernels 1, 2, and 3 have both a PL part and an AIE part, while Kernel 4 only has a PL part. 
This approach does not use many of the connectivity features of the AIE that might, for example, enable direct kernel to kernel communication staying within the AIE array.
We choose a more pragmatic approach initially to achieve functionality first before exploring the other features that should provide more efficient implementations.

With the Galapagos framework modified for Versal, we can place kernels on multiple Versal devices. With a few modifications to the Cluster Builder and allowing it to support building kernels that have both PL and AIE parts, we can use the same tool flow to build multiple Galapagos Clusters on Versal ACAP devices.

\subsection{Estimating Performance for I-BERT on Multiple Versal ACAP Devices}\label{CH-VERSAL-est}

\begin{figure}
    \centerline{
    \includegraphics[width=1\linewidth]{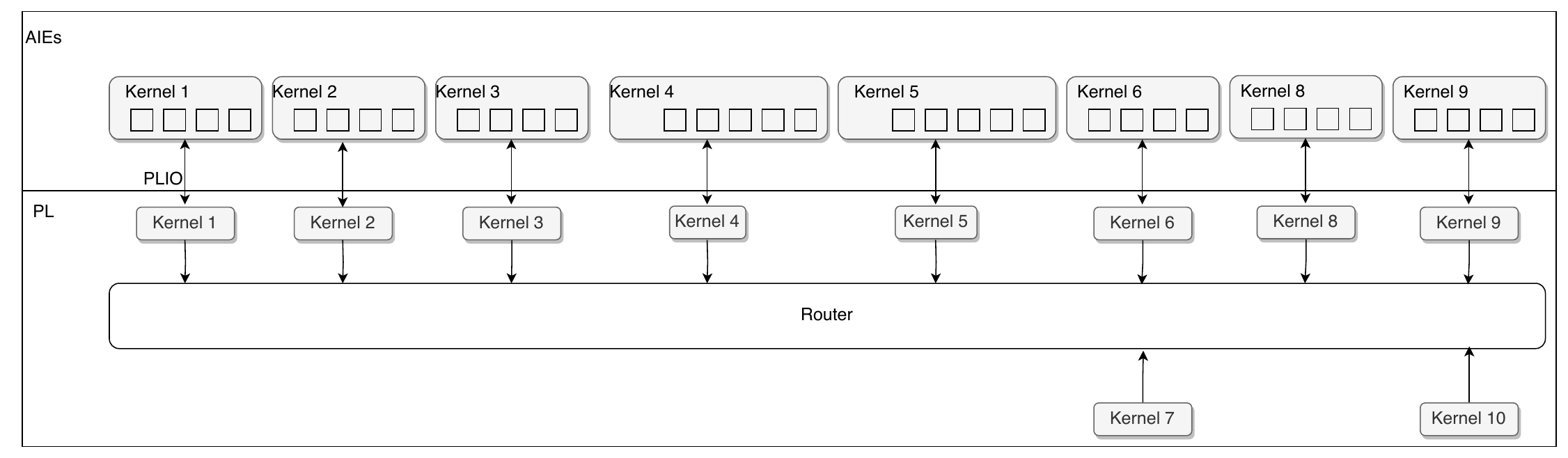}
    }
    \caption{Architecture of the I-BERT Encoder Layer on Versal ACAP}
    \label{fig:versal-encoder}
\end{figure}

Based on the modifications to Galapagos, we can now map the encoders of I-BERT to Galapagos kernels. We assume all the input sequences are padded to the maximum sequence length of 128 for ease of estimation. Figure~\ref{fig:versal-encoder} shows the mapping of one encoder to one VCK190 board. 

Kernels 1, 2, 3, and 6 on the AIE side are Matrix-Multiply kernels with dimensions of $128\times 768 \times 768$. To fully utilize the AIE, the weight matrix needs to be stored in the data memory (each AIE has one data memory of 32KB), and the input matrix should be transferred through the network and stored in the register file (2KB). This is because the AIE processor can access the weight at each clock cycle without the bottleneck of transferring data from PL to AIEs and transferring data from DRAM to PL. To store a weight matrix of size 768x768, we need 576 KB of memory for INT8 data format, and each AIE has 32 KB of data memory, thus we need at least 18 AIEs. We choose to partition the matrix into 24 sub-matrices with a dimension of $768\times 32$ and use 24 AIEs. Other configurations can also be considered here, for example, we can partition the matrix into a grid of $3\times 8$ partial matrices with a dimension of $256\times 96$, and utilize a grid of $3\times 8$ AIEs for each of the Kernels 1, 2, 3, and 6.

\begin{figure}
    \centerline{
    \includegraphics[width=0.6\linewidth]{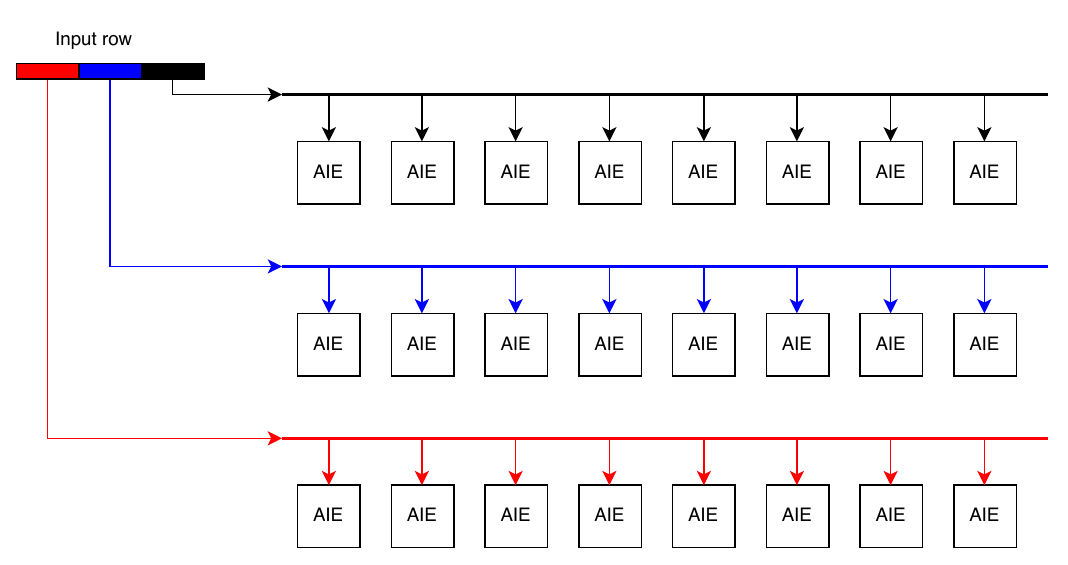}
    }
    \caption{Matrix-Multiply on a grid of $3\times 8$ AIEs}
    \label{fig:versal-matmul}
\end{figure}

Each AIE performs a matrix-multiply for a partial weight matrix, and the results can be gathered and sent back to the PL side. The PL side of Kernels 1, 2, and 3 implements the Quant modules. The input matrix can be transferred in a row-wise fashion, the same as the current design, to each of the 24 AIEs. Figure~\ref{fig:versal-matmul} shows the configuration where the matrix is partitioned into a grid of $3\times 8$ AIEs. The input row of the matrix is split into three segments and each segment is sent to a different row using the packet-switched connection. Each segment is broadcast to eight AIEs in the same row using the broadcast chain. 

Kernel 4 implements 12 Attention Dot-Product modules and 12 Softmax modules. This is different than the current design, where we implement them in separate kernels. This is due to the consideration that there are only 39 PLIOs connecting the PL and the AIEs, thus it is important to limit the number of kernels in one Versal device to save the PLIOs. For each Attention Dot-Product, the matrix-multiply dimension is $128\times 64\times 128$. The Softmax can also be implemented on the PL side. Kernel 4 is assigned 12 AIEs so that each Dot-Product Attention module can occupy one AIE. Kernel 5 implements 12 Softmax Matrix-Multiply modules. Each matrix-multiply has a dimension of $128\times 128\times 64$. Kernel 5 is also assigned 12 AIEs. Kernels 7 and 10 implement the LayerNorm modules, which can be implemented only on the PL side. Kernels 8 and 9 on the AIE side implement the Matrix-Multiply module and have the same dimension of $128\times 768 \times 3072$. Since the number of operations is four times the number of operations in Kernel 1, 2, 3, and 6, we assign 96 AIEs to each of the Kernels 8 and 9, so that the throughput can match Kernels 1, 2, 3, and 6.
Overall, there are $24\times 4 + 12 + 12 + 96\times 2 = 312$ AIEs for one encoder layer.

Because each AIE can fetch 512 bits of data, or 64 8-bit weights, per cycle from the data memory, and the 64 weights are used to perform dot-product with the input partial row, we use 64 multiplies per cycle to estimate the latency of each kernel. For Kernels 1, 2, 3, and 6, there are $128\times 768\times 32 = 3,145,728$ multiplications for each AIE, then the number of clock cycles is $3,145,728 \mathbin{/} 64 = 49,152$. The AIEs are set to operate on 1GHz, so the latency is $49,152  \mathbin{/} 10^9 = 49 \mu s$. For Kernels 4 and 5, there are $128\times 128\times 64 = 1,048,576$ operations for each of the Matrix Multiply, thus the latency is $16 \mu s$. For Kernels 8 and 9, the latency is also $49 \mu s$.

Our approach for this estimation was discussed with an engineer at AMD and deemed to be reasonable~\cite{AMDvalidation1}.
The engineer also suggested that with more analysis there might even be another factor of 2 performance available with more careful data placement in the memories.
Another AMD engineer~\cite{AMDvalidation2} pointed out that the AIE-ML~\cite{aie-ml} version of the AIE is designed to do matrix multiplications better than the original AIE that we are using.
In our analysis, we will use our original estimate, but there could be even more performance available with the Versal architecture that can only be explored with an actual implementation.

To calculate the network latency overhead, we assume the 12 Versal devices are connected to the same 100G switch. We use the latency measured between two FPGAs, which is about $1.1 \mu s$. Quant, GELU, Softmax, and LayerNorm add a latency overhead of $26.1 \mu s$, thus the overall latency for one encoder is $98 + 26.1 = 124.1 \mu s$.  Based on Equation~\ref{eqn:est-latency}, when the sequence length is 128, $X$ is about 0.53 times of $T$. We set $T$ equal to $124.1 \mu s$, $L$ equal to 12, $X$ equal to $65.8 \mu s$, and $d$ equal to $1.1 \mu s$. After adding network latency and nonlinear function latency, the estimated overall latency for I-BERT is $860 \mu s$.

According to \cite{nv-bert}, the NVIDIA A100 GPU~\cite{nv-a100} has an average batch-1 latency of $770 \mu s$ for the BERT base model inference with a maximum sequence length of 128 and INT8 precision compared to our estimate of $860 \mu s$ realizing it is possible that there are opportunities to significantly improve the Versal latency. The A100 GPU can achieve up to 1248 TOPs of throughput for INT8, while the Versal VCK190 can only achieve up to 133 TOPs for INT8, which is only about 9.4\% of the A100 throughput. There is likely a power efficiency benefit, but that analysis is left for future work. 
The main observation from these results is that the Versal latency estimate is comparable to the high-end GPUs.

It is also important  to point out that our estimate uses 12 Versal cards.  
If we want to compare a single FPGA implementation to a single GPU implementation mainly to argue that FPGAs can compete with GPUs, we need to make some adjustments to our architecture.
It should be possible to reconfigure the weights in a card after its processing stage is complete so two cards would suffice.
One card will be computing and the other will be reconfigured with new weights.
Since communication between cards is over the network, it is straightforward to direct the output of one card to the appropriate input of another card.
With more detailed analysis it should be possible to reconfigure weights while processing is occurring within one card so that only one card is required.
This is strong evidence that a single Versal device is competitive with a single GPU in performance, but realize that our goal is build multi-FPGA applications to compete with multi-GPU applications, so with this particular analysis our only goal is to show that a modern FPGA can compete with a modern GPU for this application.

According to our estimates, a Versal implementation is comparable to a high-end GPU for I-BERT when considering batch-1 latencies, which would be used for low-latency applications.
The main conclusion of the comparative analysis is that a Versal FPGA is a viable device for building performance-competitive implementations of transformers when compared to GPUs and there is value to continue exploring multi-FPGA implementations of applications like LLMs because they should be competitive with clusters of GPUs.
More importantly, we would expect there to be a reasonable power advantage given past experiences of comparing applications implemented on FPGAs and GPUs.
This can be validated after an actual Versal implementation is completed but we have high confidence of a favourable result.
The implementations of transformer-based models done in FlightLLM~\cite{flightllm:fpga2024} and by Zhuang et al.~\cite{zhou:fpga2024} on Versal devices have already shown better performance and energy efficiency than high-end GPUs.

\subsection{Scalability}

I-BERT is small enough to fit on one GPU, but scaling to larger transformers requires many more GPUs.
GPU scalability when the GPUs are attached as accelerators to CPUs is not as easy as with Galapagos because inter-GPU communication over the data center network requires a connection through a PCIe interface to a network port, which has a high overhead.
One way to achieve more efficient communication in a cluster of GPUs is to use an interconnect such as NvLink and NvSwitch~\cite{nvlink}, but that is proprietary to only one GPU vendor.

For our architecture, scaling just means adding more FPGAs onto the network and building larger applications.
We are confident that Galapagos can continue to scale as we have built larger systems to the extent of the resources we have available.
In our proof-of-concept I-BERT encoder, we have 38 kernels, including six GMI kernels spread over six FPGAs.
We have built systems with up to 96 kernels as microbenchmarks across six FPGAs~\cite{galapagos}.
In AIgean~\cite{aigean} we have built nine and 12 FPGA systems with over 100 kernels.
While these are still small numbers, the architecture of our system is much like what is used in Catapult~v2~\cite{catapult_v2}, which is an existence proof that large network-connected FPGA platforms can be built.
Galapagos itself only builds the FPGAs and its limit is the 256 kernels in a cluster that we address by building clusters of clusters to enable better scaling as discussed in Section~\ref{CH-SCALING}.

The other consideration is the overhead of Galapagos and how it affects communication latencies.
In~\cite{aigean} measurements showed that with our 100~Gbps UDP we were able to achieve full bandwidth on throughput at a FPGA-to-FPGA round-trip latency of 0.17$\mu s$ through a 100~Gbps switch.
This can be compared to Catapult~v2~\cite{catapult_v2}, which reported a round-trip latency of 2.88$\mu s$ using LTL on a 40~Gbps link with a single top-of-rack switch in the middle.
While these numbers are not directly comparable, the numbers give confidence that communication is not incurring significant overhead due to Galapagos.

\section{Conclusion}\label{CH-CONCLUSION}

In this work, we explore the feasibility of building large machine learning applications, such as LLMs, using clusters of FPGAs.
We have built a working proof-of-concept implementation of a small transformer network using a cluster of six relatively small and older FPGAs.
The proof-of-concept demonstrates that our enhanced Galapagos environment, including both the hardware infrastructure and the Cluster Builder tool, has the basic functionality required to build transformers on FPGA clusters.
We have used the proof-of-concept as the basis for estimating performance using the most recent Versal FPGAs from AMD, which uses technology more comparable to the recent high-end GPUs.
With a performance model based on understanding the behavior of the proof-of-concept, we have shown that for low-latency batch-1 performance the Versal FPGAs are performance competitive with recent GPUs.
This demonstrates that building large applications with clusters of modern FPGAs shows significant promise and further implementations should be explored.
The main expected benefit is that there will be significant power efficiencies from using FPGAs while still being competitive with GPUs in performance.
Only actual implementations of systems where power can be measured can prove this hypothesis.

With our working proof-of-concept demonstration and analysis of that platform we firmly believe that it is feasible to build large machine learning applications with clusters of FPGAs and continued efforts to achieve this goal are merited.

\section{Future Work}\label{CH-FUTURE}

The promise of our work suggests many directions for future work.  We list some key next steps here:

\begin{itemize}
    \item Modify the Galapagos framework so that it can support the Versal architecture. As mentioned in Section~\ref{CH-VERSAL}, this requires the modified Galapagos to leverage the NoC and AIEs in the Versal devices. The modified Galapagos kernels should be able to utilize both the AIEs and the PL.
    \item After modifying Galapagos for Versal, the most important next task is to build a working Versal implementation of I-BERT and analyse its power usage.
    This can then be compared to GPU implementations, which
    will provide a better idea about the energy efficiency of FPGAs versus GPUs for low-latency transformer implementations.
    If FPGAs are shown to have a significant advantage based on this comparison then the expectation is the trend would continue when scaling to multiple FPGAs.
    This would be a strong argument to continue exploring multi-FPGA implementations of large-scale machine learning applications.
    \item Examine opportunities to improve the implementation of I-BERT on Versal as suggested by the AMD engineers who reviewed our design.
    \item Investigate more efficient implementations of the various components of the enhanced Galapagos that was used to build the current proof-of-concept.

    \item Design a runtime agent for Galapagos Clusters. The runtime agent is essential when the number of FPGAs is not enough for deploying all the Galapagos Clusters at initialization, and we need a runtime agent to dynamically deploy and swap clusters. 
    This will be especially useful when prototyping large systems with a limited amount of hardware.

    \item Begin the exploration of accelerating larger transformers, such as GPT-3~\cite{gpt-3}, on multiple Versal devices. It might take dozens or even hundreds of Versal devices to accelerate one decoder layer in models like GPT-3, thus it is essential to have a runtime agent to perform parameter updates for different decoder layers, as a means to reduce the total number of Versal Devices required.
    An important consideration for larger transformers is the amount of memory required so the memory requirements will need to be understood before designing an appropriate memory system.
    
\end{itemize}

\begin{acks}
This work is supported by NSERC and AMD/Xilinx.
\end{acks}

\bibliographystyle{ACM-Reference-Format}
\bibliography{main}

\end{document}